\documentclass[11pt]{article}

\usepackage{amsmath}
\usepackage{amssymb}

\usepackage{graphicx}
\usepackage{epstopdf}

\usepackage{cite}

\usepackage{color} 

\usepackage[labelfont=bf,labelsep=period,justification=raggedright]{caption}

\makeatletter
\renewcommand{\@biblabel}[1]{\quad#1.}
\makeatother

\newcommand{\revs}[1] {#1}

\date{}

\begin{document}

\begin{flushleft}
{\Large
\textbf{Dynamic effective connectivity of inter-areal brain circuits}
}

Demian Battaglia$^{1, 3\ast}$, 
Annette Witt$^{1, 2, 3}$,
Fred Wolf$^{1, 3}$,
Theo Geisel$^{1, 3}$
\\
\bf{1} Max Planck Institute for Dynamics and Self-Organization, G\"ottingen, Germany
\\
\bf{2} German Primate Center, G\"ottingen, Germany
\\
\bf{3} Bernstein Center for Computational Neuroscience, G\"ottingen, Germany
\\
$\ast$ E-mail: demian@nld.ds.mpg.de 
\end{flushleft}

\section*{Abstract}

Anatomic connections between brain areas 
affect information flow between neuronal circuits and the synchronization of neuronal activity.
However, such structural connectivity does not coincide with effective 
connectivity, related to the more elusive question ``Which areas cause 
the present activity of which others?''. Effective connectivity is directed and depends flexibly on contexts and tasks. Here we show that a dynamic effective connectivity can emerge from
transitions in the collective organization of coherent neural activity. 
Integrating simulation and semi-analytic approaches, we study mesoscale network motifs of interacting cortical areas, modeled as large random networks of spiking neurons or as simple rate units. Through a causal analysis of time-series of model neural activity, we show that different dynamical states generated by a same structural connectivity motif correspond to distinct effective connectivity motifs. Such effective motifs can display a dominant directionality, due to spontaneous symmetry breaking and effective entrainment between local brain rhythms, although all connections in the considered structural motifs are reciprocal.
We show then that transitions between effective connectivity configurations (like, for instance, reversal in the direction of inter-areal interactions) can be triggered reliably by brief perturbation inputs, properly timed with respect to an ongoing local oscillation, without the need for plastic synaptic changes.
Finally, we analyze how the information encoded in spiking patterns of a local neuronal population is propagated across a fixed structural connectivity motif, demonstrating that changes in the active effective connectivity regulate both the efficiency and the directionality of information transfer.
Previous studies stressed the role played by coherent oscillations in establishing efficient communication between distant areas. Going beyond these early proposals, we advance here that dynamic interactions between brain rhythms provide as well the basis for the self-organized control of this ``communication-through-coherence'', making thus possible a fast ``on-demand'' reconfiguration of global information routing modalities.

\section*{Author Summary}

The circuits of the brain must perform a daunting amount of functions. But how can ``brain states'' be flexibly controlled, given that anatomic inter-areal connections can be considered as fixed, on timescales relevant for behavior? We hypothesize that, thanks to the nonlinear interaction between brain rhythms, even a simple circuit involving few brain areas can originate a multitude of effective circuits, associated with alternative functions selectable ``on demand''.
A distinction is usually made between structural connectivity, which describes actual synaptic connections, and effective connectivity, quantifying, beyond correlation, directed inter-areal causal influences. In our study, we measure effective connectivity based on time-series of neural activity generated by model inter-areal circuits. We find that ``causality follows dynamics''. We show indeed that different effective networks correspond to different dynamical states associated to a same structural network (in particular, different phase-locking patterns between local neuronal oscillations). We then find that ``information follows causality'' (and thus, again, dynamics). We demonstrate that different effective networks give rise to alternative modalities of information routing between brain areas wired together in a fixed structural network. Thus, \revs{we propose that the flow of ``digital-like'' information encoded in spiking patterns is controlled by the self-organization of interacting ``analog'' rate oscillations.}

\section*{Introduction}

\begin{figure}[!ht]
\begin{center}
\includegraphics[width=8.3cm]{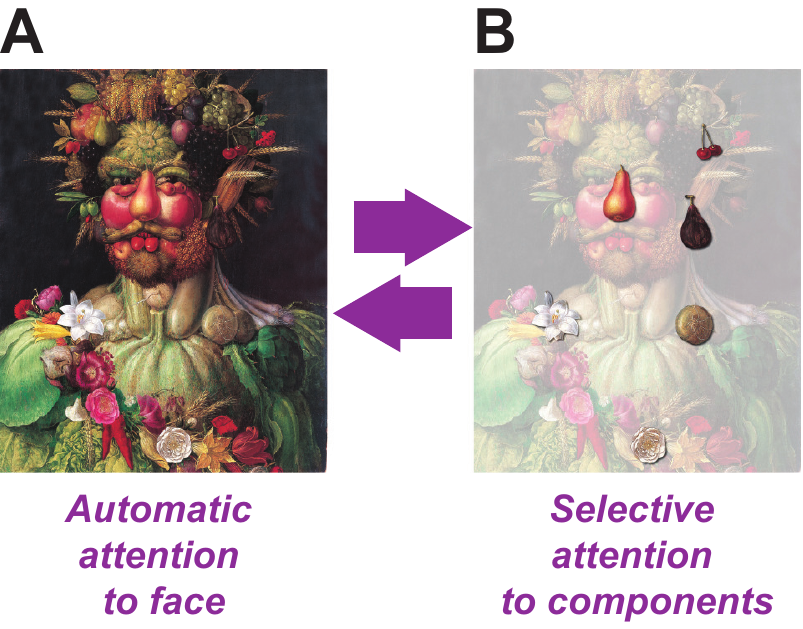} 
\end{center}
\caption{ }
\label{fig:ARCIMBOLDO}
\end{figure}

In Arcimboldo's (1527-1593) paintings, whimsical portraits emerge out of arrangements of flowers and vegetables. Only directing attention to details, the illusion of seeing a face is suppressed (Fig.~\ref{fig:ARCIMBOLDO}A-B). Our brain is indeed hardwired to detect facial features and a complex network of brain areas is devoted to face perception \cite{Fairhall2007}. The capacity to detect faces in an Arcimboldo canvas may be lost when lesions impair the connectivity between these areas \cite{Steeves2006}. It is not conceivable, however, that, in a healthy subject, shifts between alternate perceptions are obtained by actual ``plugging and unplugging'' of synapses, as in a manual telephone switchboard. 

Brain functions ---from vision \cite{Vuilleumier2007} or motor preparation \cite{Brovelli2004} up to memory \cite{Clapp2011}, attention \cite{Rossi2009, Corbetta2002, Zanto2011} or awareness \cite{Gaillard2009}--- as well as their complex coordination \cite{Tononi1998} require the control of inter-areal interactions on time-scales faster than synaptic changes \cite{Bressler2001, Varela2001}. In particular, strength and direction of causal influences between areas ---described by the so-called {\it effective} connectivity \revs{(or, in a more restrictive sense, \textit{causal} connectivity) \cite{Friston1994, Ioannides2007, Friston2011, Bressler2011}}--- must be reconfigurable even when the underlying {\it structural} (i.e.~anatomic) connectivity is fixed. The ability to quickly reshape effective connectivity is a chief requirement for performance in a changing environment. Yet it is an open problem  to understand which circuit mechanisms allow for achieving this ability. How can manifold effective connectivities ---corresponding to different patterns of inter-areal interactions, or brain states \cite{Gilbert2007}--- result from a fixed structural connectivity? And how can effective connectivity be controlled without resorting to structural plasticity, leading to a flexible ``on demand'' selection of function? 

\revs{Several experimental and theoretical studies have suggested that \textit{multi-stability} of neural circuits might underlie the switching between different perceptions or behaviors \cite{Haken1985, Ditzinger1989, Lumer1998, Deco2007, MorenoBote2007}. In this view, transitions between many possible attractors of the neural dynamics would occur under the combined influence of structured ``brain noise'' \cite{Misic2010} and of the bias exerted by sensory or cognitive driving \cite{Deco2008, Deco2009, Deco2011}. Recent reports have more specifically highlighted how dynamic multi-stability can give rise to transitions between different oscillatory states of brain dynamics \cite{Freyer2009, Freyer2011}. This is particularly relevant in this context, because long-range oscillatory coherence \cite{Varela2001, Wang2010} ---in particular in the gamma band of frequency (30-100 Hz) \cite{Wang2010, Engel2001, Fries2005, Fries2007}--- is believed to play a central role in inter-areal communication.} 

\revs{Ongoing local oscillatory activity modulates rhythmically neuronal excitability \cite{Volgushev1998}. As a consequence, according to the influential \textit{communication-through-coherence} hypothesis \cite{Fries2005}, neuronal groups oscillating in a suitable phase coherence relation ---such to align their respective ``communication windows''--- are likely to interact more efficiently than neuronal groups which are not synchronized. However, despite accumulating experimental evidence of communication-through-coherence mechanisms \cite{Schoffelen2005, Womelsdorf2007, Ghazanfar2008, Canolty2010, Hipp2011} and of their involvement in selective attention and top-down modulation \cite{Engel2001, Fries2008, Gregoriou2009}, a complete understanding of how inter-areal phase coherence can be flexibly regulated at the circuit level is still missing. In this study we go beyond earlier contributions, by showing that the self-organization properties of interacting brain rhythms lead spontaneously to the emergence of mechanisms for the robust and reliable control of inter-areal phase-relations and information routing.}

\revs{Through large-scale simulations of networks of spiking neurons and rigorous analysis of mean-field rate models, we model the oscillatory dynamics of generic brain circuits involving a small number of interacting areas (\textit{structural connectivity motifs} at the mesoscopic scale). Following \cite{Honey2007}, we extract then the effective connectivity associated to this simulated neural activity. In the framework of this study, we use a data driven rather than a model driven approach to effective connectivity \cite{Bressler2011} (see also {\it Discussion} section), and we quantify causal influences in an operational sense, based on a statistical analysis of multivariate time-series of synthetic ``LFP'' signals. Our causality measure of choice is} Transfer Entropy (TE) \cite{Schreiber2000, Kaiser2002}. TE is based on information theory  \cite{MacKay2003} (and therefore more general than causality measures based on regression \cite{Granger1969, Ding2006}), is ``model-agnostic'' and in principle capable of capturing arbitrary \revs{linear and nonlinear} inter-areal interactions. 

\revs{Through our analyses,  we first confirm the intuition that} ``causality follows dynamics''. \revs{Indeed we show that our causal analysis based on TE is able to capture the complex multi-stable dynamics of the simulated neural activity. As a result, different \textit{effective connectivity motifs} stem out of different dynamical states of the underlying structural connectivity motif (more specifically, different phase-locking patterns of coherent gamma oscillations). Transitions between these effective connectivity motifs correspond to switchings between alternative dynamic attractors.}

\revs{We show then that transitions can be} reliably induced through brief transient perturbations properly timed with respect to the ongoing rhythms\revs{, due to the non-linear phase-response properties \cite{Kuramoto1984} of oscillating neuronal populations.} Based on dynamics, this neurally-plausible mechanism for brain-state switching is metabolically more efficient than coordinated plastic changes of a large number of synapses, and is faster than neuromodulation \cite{Constantinople2011}.

\revs{Finally, we} find that ``information follows causality'' (and, thus, again, dynamics). \revs{As a matter of fact, effective connectivity is measured in terms of time-series of ``LFP-like'' signals reflecting collective activity of population of neurons, while the information encoded in neuronal representations is carried by spiking activity. Therefore an effective connectivity analysis ---even when based on TE--- does not provide an actual description of information transmission in the sense of neural information processing and complementary analyses are required to investigate this aspect.} Based on a general information theoretical perspective, which does not require specifying details of the used encoding \cite{MacKay2003}, we \revs{consider} information encoded in spiking patterns \cite{MacKay1952, Mainen1995, Reich2001, Osborne2008, Ohiorhenuan2010}, rather than in modulations of the population firing rate. As a matter of fact, the spiking of individual neurons can be very irregular even when the collective rate oscillations are regular \cite{Brunel1999, Brunel2003, Brunel2006, Brunel2008}. Therefore, even local rhythms in which the firing rate is modulated in a very stereotyped way, might correspond to irregular (highly entropic) sequences of codewords encoding information in a digital-like fashion (e.g. by the firing ---``1''---  or missed firing ---``0''--- of specific spikes at a given cycle \cite{Strong1998}). In such a framework, oscillations would not directly represent information, but would rather act as a carrier of ``data-packets'' associated to spike patterns of synchronously active cell assemblies. \revs{By quantifying} through a Mutual Information (MI) analysis {the maximum amount of information encoded potentially in the spiking activity of a local area and by evaluating how much of this information is actually transferred to distant interconnected areas, we demonstrate} that different effective connectivity configurations correspond to different modalities of information routing. Therefore, the pathways along which information propagates can be reconfigured within the time of a few reference oscillation cycles, by switching to a different effective connectivity motif. 

Our results provide thus novel theoretical support to the hypothesis that dynamic effective connectivity stems from the self-organization of brain rhythmic activity. Going beyond previous proposals, which stressed the importance of oscillations for feature binding \cite{NeuronBindingIssue1999} or for efficient inter-areal ``communication-through-coherence'', we advance that the complex dynamics of interacting brain rhythms allow to implement reconfigurable routing of information \revs{in a self-organized manner and }in a way reminiscent of a clocked device (in which digital-like spike pattern codewords are exchanged at each cycle of an analog rate oscillation).


\section*{Results}

\subsection*{Models of interacting areas}

\begin{figure}[!ht]
\begin{center}
\includegraphics[width=8.3cm]{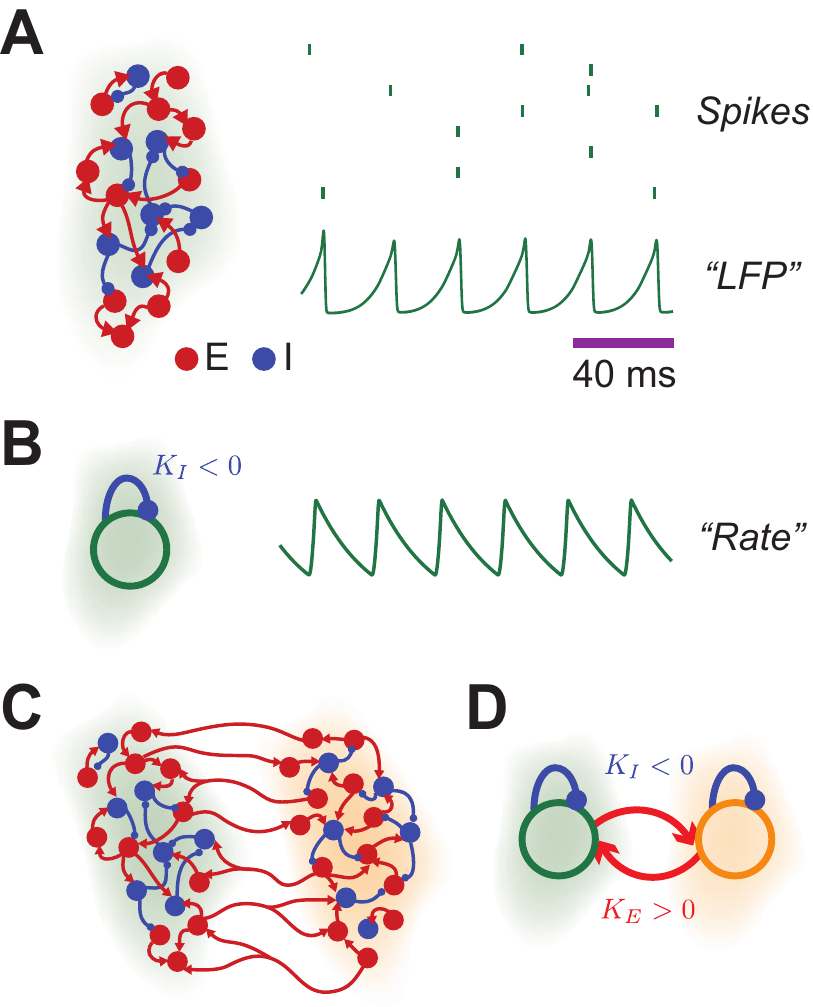} 
\end{center}
\caption{ }
\label{fig:MODELS}
\end{figure}

In order to model the neuronal activity of interacting areas, we use two different approaches, previously introduced in \cite{Battaglia2007}. First, each area is modeled as a large network of \revs{thousands} of excitatory and inhibitory spiking neurons, driven by uncorrelated noise representing background cortical input ({\it network model}). Recurrent synaptic connections are random and sparse. In these networks, local interactions are excitatory and inhibitory. A scheme of the network model for a local area is depicted in Fig.~\ref{fig:MODELS}A (left). In agreement with experimental evidence that the recruitment of local interneuronal networks is necessary for obtaining coherent gamma cortical activity {\it in vitro} and {\it in vivo} \cite{Bartos2007, Cardin2009},  the model develops synchronous oscillations ($\sim 50$ Hz) when inhibition is strong, i.e.~for a sufficiently large probability $p_I$ of inhibitory connection \cite{Brunel1999, Brunel2003, Brunel2006, Brunel2008, Whittington1995}. These fast oscillations are clearly visible in the average membrane potential (denoted in the following as ``LFP''), an example trace of which is represented in Fig.~\ref{fig:MODELS}A (bottom right). Despite the regularity of these collective rhythms, the ongoing neural activity is only sparsely synchronized. The spiking of individual neurons is indeed very irregular \cite{Brunel1999, Brunel2006} and neurons do not fire an action potential at every oscillation cycle, as visible from the example spike trains represented in Fig.~\ref{fig:MODELS}A (top right). Structural network motifs involving $N \ge 2$ areas are constructed by allowing excitatory neurons to establish in addition long-range connections toward excitatory or inhibitory neurons in a distant target area (see a schematic representation of an $N=2$ structural connectivity motif in Fig.~\ref{fig:MODELS}C). The strength of inter-areal coupling is regulated by varying the probability $p_E$ of establishing an excitatory connection.

In a second analytically more tractable approach, each area is described by a mean-field firing rate variable ({\it rate model}). The firing rate of a local population of neurons obeys the non-linear dynamical equation \eqref{eq:ratemodel} (see {\it Methods}). All incorporated interactions are delayed, accounting for axonal propagation and synaptic integration. Local interactions are dominantly inhibitory (with coupling strength $K_I < 0$ and delay $D$).  Driving is provided by a constant external current. A cartoon of the rate model for a local area is depicted in Fig.~\ref{fig:MODELS}B (left). As in the network model, the firing rates undergo fast oscillations for strong inhibition ($K_I < K_I^{(c)} \simeq -\frac{\pi}{2D}$, \cite{Battaglia2007}). An example firing rate trace is shown  in Fig.~\ref{fig:MODELS}B (right). In order to build structural networks involving $N \ge 2$ areas, different mean-field units are coupled together reciprocally by excitatory long range interactions with strength $K_E > 0$ and delay $\overline{D} \geq D$ (see a schematic representation of an $N=2$ structural motif in Fig.~\ref{fig:MODELS}D). Remarkably, the rate model and the network model display matching dynamical states \cite{Battaglia2007} (see also later, Figures~\ref{fig:EFFECTI_unidir},~\ref{fig:EFFECTI_leaky}~and\ref{fig:EFFECTI_mutual}). More details on the network and the rate models are given in the {\it Methods} section and in the Supporting~Text~S1.

\subsection*{Causality follows dynamics} 

For simplicity, we study fully connected structural motifs involving a few areas ($N=2,3$). \revs{Note however that our approach might be extended to other structural motifs \cite{Sporns2004} or even to larger-scale networks with more specific topologies \cite{Honey2007, Breakspear2010}.} 

\revs{In the simple structural motifs we consider, delays and strengths of local excitation and inhibition are homogeneous across different areas. Long-range inter-areal connections are as well isotropic, i.e. strengths and delays of inter-areal interactions are the same in all directions. Delay and strength of local and long-range connections can be changed parametrically, but only in a matching way for homologous connections, in such a way that the overall topology of the structural motif is left unchanged.} As previously shown in \cite{Battaglia2007}, \revs{different dynamical states ---characterized by oscillations with different phase-locking relations and degrees of periodicity--- can arise from these simple structural motif topologies.  Changes in the strength of local inhibition, of long-range excitation or of delays of local and long-range connections can lead to phase transitions between qualitatively distinct dynamical states. Interestingly, however, within broad ranges of parameters, multi-stabilities between dynamical states with different phase-locking patterns take place even for completely fixed interaction strengths and delays.}

\revs{We generate} multivariate time-series of simulated ``LFPs'' in \revs{different dynamical states of our models and we calculate TEs for all the possible directed pairwise interactions. We show then that effective connectivities associated to different dynamical states are also different. The resulting effective connectivities can be} depicted in diagrammatic form \revs{by drawing an arrow for each statistically significant causal interaction. The thickness of each arrow encodes the strength of the corresponding interaction. This graphical representation makes apparent, then, that }effective connectivity motifs or, more briefly, {\it effective motifs}, \revs{with many different topologies emerge from structural motifs with a same fixed topology. Such effective motifs are organized into \textit{families}. All the motifs within a same family correspond to dynamical states which are multi-stable for a given choice of parameters, while different families of motifs are obtained for different ranges of parameters leading to different ensembles of dynamical states.} 

\begin{figure}[!ht]
\begin{center}
\includegraphics[width=12.35cm]{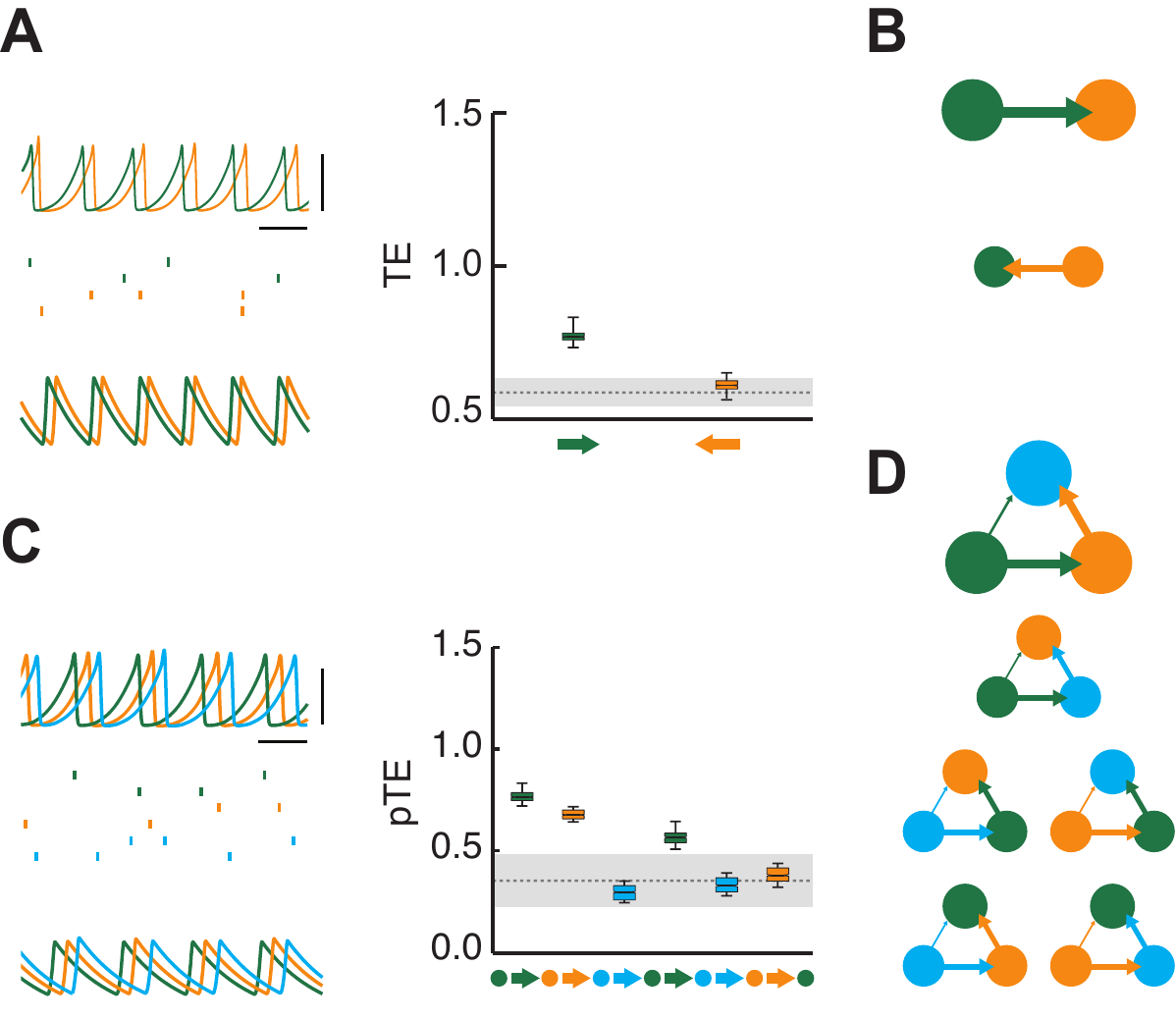} 
\end{center}
\caption{ }
\label{fig:EFFECTI_unidir}
\end{figure}

\begin{figure}[!ht]
\begin{center}
\includegraphics[width=12.35cm]{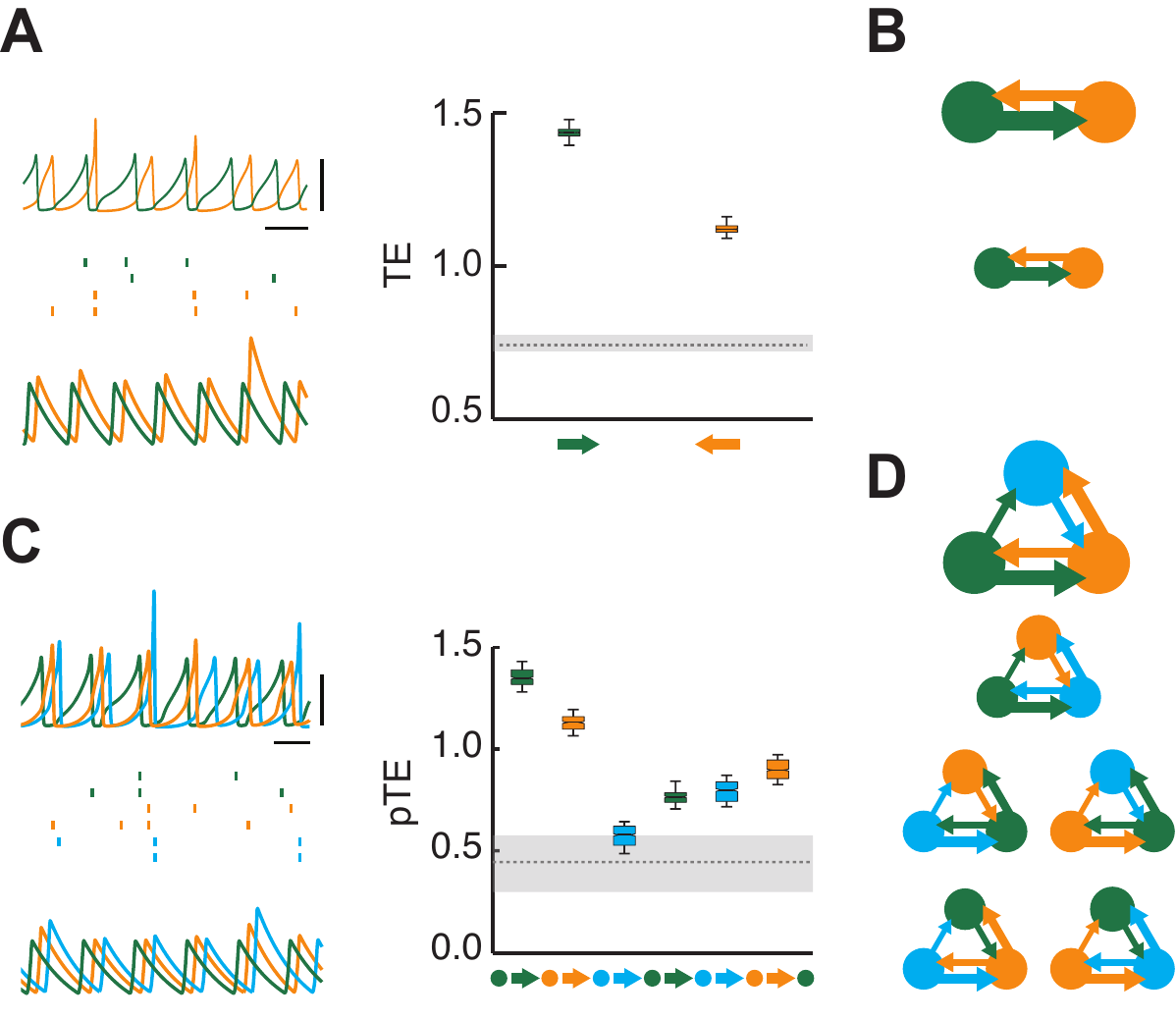} 
\end{center}
\caption{ }
\label{fig:EFFECTI_leaky}
\end{figure}

\begin{figure}[!ht]
\begin{center}
\includegraphics[width=12.35cm]{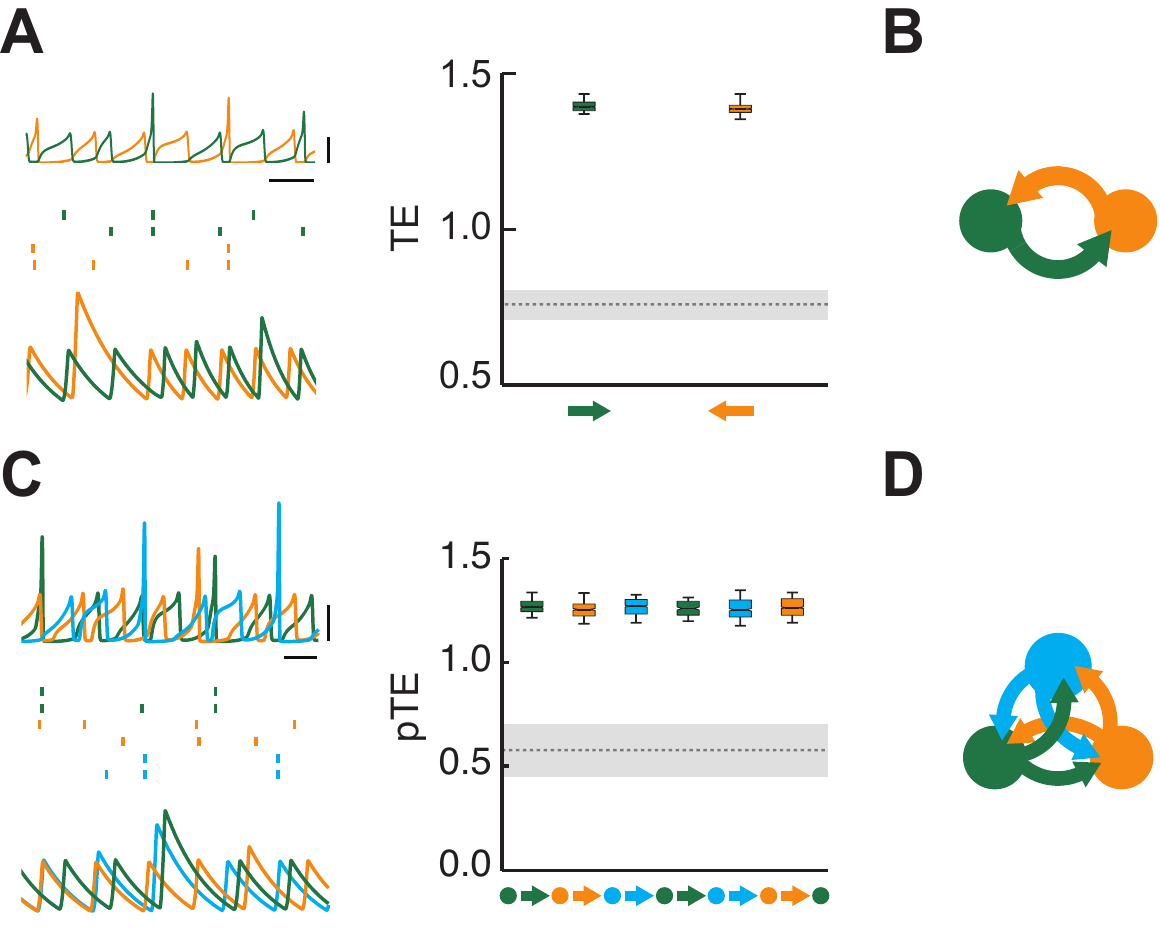} 
\end{center}
\caption{ }
\label{fig:EFFECTI_mutual}
\end{figure}

\revs{We analyze in detail, in Figures~\ref{fig:EFFECTI_unidir},~\ref{fig:EFFECTI_leaky}~and~\ref{fig:EFFECTI_mutual}, three families of motifs arising for strong intra-areal inhibition and similarly small values of delays for local and long-range connections.  We consider $N=2$ (panels A and B) and $N=3$ (panels C and D) structural motifs. Panels A and C show TEs for different directions of interaction, together with ``LFPs''  and example spike trains (from the network model), and rate traces (from matching dynamical states of the rate model). Panels B and D display motifs belonging to the corresponding effective motif families.} 

\revs{A first family of effective motifs occurs for weak inter-areal coupling}. In this case, neuronal activity oscillates in a \revs{roughly} periodic fashion (Fig.~\ref{fig:EFFECTI_unidir}A~and~C, left sub-panel). When local inhibition is strong, \revs{the local oscillations generated within different areas lock in an out-of-phase fashion}. It is therefore possible to identify a {\it leader} area whose oscillations lead in phase over the oscillation of {\it laggard} areas \cite{Battaglia2007}. \revs{In this family, causal interactions are statistically significant only for pairwise interactions proceeding from a phase-leading area to a phase-lagging area, as shown by the the box-plots of Fig.~\ref{fig:EFFECTI_unidir}A and C (right sub-panel, see {\it Discussion} and {\it Methods} for a discussion of the threshold used for statistical significancy). As commented more in detail in the {\it Discussion} section, the anisotropy of causal influences in leader-to-laggard and laggard-to-leader directions can be understood in terms of the communication-through-coherence theory. Indeed } the longer latency from the oscillations of the laggard area to the oscillations of the leader area \revs{reduces the likelihood that rate fluctuations originated locally within a laggard area trigger correlated rate fluctuations within a leading area \cite{Womelsdorf2007} (see also {\it Discussion}). } Thus, out-of-phase lockings for weak inter-areal coupling give rise to a family of {\it unidirectional driving} effective motifs. \revs{In the case of $N=2$, causality is significant only in one of two possible directions (Fig.~\ref{fig:EFFECTI_unidir}B), depending on which of the two areas assumes the role of leader. In the case of $N=3$, } it is possible to identify a ``causal source'' area and a ``causal sink'' area (see \cite{Seth2005} for an analogous terminology), such that no direct or indirect causal interactions in a backward sense from the sink area to the source area are statistically significant. \revs{Therefore, the unidirectional driving effective motif family for $N=3$ contains six motifs (Fig.~\ref{fig:EFFECTI_unidir}D), corresponding to all the possible combinations of source and sink areas.}

\revs{A second family of effective motifs occurs for intermediate inter-areal coupling. In this case, the periodicity of the ``LFP'' oscillations is disrupted by the emergence of large correlated fluctuations in oscillation cycle amplitudes and durations. As a result, }  the phase-locking between ``LFPs''  \revs{becomes} only approximate, \revs{even if it} continues to be out-of-phase on average.  The rhythm of the laggard area is now more irregular than the rhythm in the leader area. Laggard oscillation amplitudes and durations in fact fluctuate chaotically (Fig.~\ref{fig:EFFECTI_leaky}A~and~C, left sub-panel). Fluctuations in cycle length do occasionally shorten the laggard-to-leader latencies, enhancing non-linearly and transiently the influence of laggard areas on the leader activity. 
Correspondingly, \revs{TEs in leader-to-laggard directions continue to be larger, but TEs in laggard-to-leader directions are now also statistically significant (Fig.~\ref{fig:EFFECTI_leaky}A~and~C, right sub-panel). The associated effective motifs are no more unidirectional, but continue to display a dominant direction or sense of rotation (Fig.~\ref{fig:EFFECTI_leaky}B~and~D). We refer to this family of effective motifs as to a family of} {\it leaky driving} effective motifs \revs{(containing two motifs for $N=2$ and six motifs for $N=3$).}

\revs{Finally, a third family of effective motifs occurs for stronger inter-areal coupling. In this case the rhythms of all the areas become equally irregular, characterized by an analogous level of fluctuations in cycle and duration amplitudes. During brief transients, leader areas can still be identified, but these transients do not lead to a stable dynamic behavior and different areas in the structural motif } continually exchange their leadership role (Fig.~\ref{fig:EFFECTI_mutual}A~and~C, left sub-panel). \revs{As a result of the instability of phase-leadership relations, only average TEs can be evaluated, yielding to equally large TE values for all pairwise directed interactions (Fig.~\ref{fig:EFFECTI_mutual}A~and~C, right sub-panel). This results in a family containing a single} {\it mutual driving} effective motif (Fig.~\ref{fig:EFFECTI_mutual}B~and~D).

\revs{Further increases of the inter-areal coupling strength do not restore stable phase-locking relations and, consequently, do not lead to additional families of effective motifs. Note however} that the effective motif families explored in Figures~\ref{fig:EFFECTI_unidir},~\ref{fig:EFFECTI_leaky}~and~\ref{fig:EFFECTI_mutual} are not the only one that can be generated by the considered fully symmetric structural motifs. Indeed other dynamical configurations exist. In particular, anti-phase locking (i.e. locking with phase-shifts of $180^\circ$ for $N=2$ and of $120^\circ$ for $N=3$) would become stable when assuming the same interaction delays and inter-areal coupling strengths of Figures~\ref{fig:EFFECTI_unidir},~\ref{fig:EFFECTI_leaky}~and~\ref{fig:EFFECTI_mutual}, but a weaker local inhibition. Assuming different interaction delays for local and long-range interactions, out-of-phase lockings continue to be very common, but in-phase and anti-phase locking can become stable even for strong local inhibition, within specific ranges of the ratio between local and long-range delays \cite{Battaglia2007}. For $N=3$, in the case of general delays, more complex combinations can arise as well, like, for instance, states in which two areas oscillate in-phase, while a third is out-of-phase. In-phase locking between areas gives rise to identical TEs for all possible directed interactions, resulting in effective motifs without a dominant directionality. Anti-phase lockings for $N=2,3$ give rise to relatively large inter-areal phase-shifts and, correspondingly, to weak inter-areal influences (at least in the case of weak inter-areal coupling), resulting in small TE levels which are not statistically significant (not shown). However, in the framework of this study, we focus exclusively on out-of-phase-locked dynamical states, because they are particularly relevant when trying to achieve a reconfigurable inter-areal routing of information (see later results and \textit{Discussion} section). 

To conclude, we remark that absolute values of TE depend on specific parameter choices (notably, on time-lag and signal quantization, see {\it Methods}). However, the relative strengths of TE in different directions ---and, therefore, the resulting topology of the associated effective motifs--- are rather robust against changes of these parameters. \revs{Robustness of causality estimation is analyzed more in detail in the {\it Discussion} section}.

\subsection*{Spontaneous symmetry breaking} 
How can asymmetric causal influences emerge from a symmetric structural connectivity? A fundamental dynamical mechanism involved in this phenomenon is known as {\it spontaneous symmetry breaking}. \revs{As shown in \cite{Battaglia2007}, for the case of the $N=2$ structural motif, a phase transition occurs at a critical value of the strength of inter-areal inhibition. When local inhibition is stronger than this critical threshold, a phase-locked configuration in which the two areas oscillate in anti-phase loses its stability in favor of a pair of out-of-phase-locking configurations, which become concomitantly stable. The considered structural motif is symmetric, since it is left unchanged after a permutation of the two areas. However, while the anti-phase-locking configuration, stable for weak local inhibition, share this permutation symmetry with the full system, this is no more true for the out-of-phase-locking configurations, stable for strong local inhibition. Note, nevertheless, that the configuration in which leader and laggard area are inverted is also a stable equilibrium, i.e. the complete set of stable equilibria continue to be symmetric, even if individual stable equilibria are not (leading thus to multi-stability). In general, one speaks about spontaneous symmetry breaking} whenever a system with specific symmetry properties assumes dynamic configurations whose degree of symmetry is reduced with respect to the full symmetry of the system. \revs{The occurrence of symmetry breaking is the signature of a phase transition (of the second order \cite{Landau1937}), which leads to the stabilization of states with reduced symmetry.}

\revs{The existence of a symmetry-breaking phase transition in the simple structural motifs we analyze here (for simplicity, we consider the $N=2$ case)} can be proven analytically for the rate model, by deriving the function $\Gamma(\Delta\phi)$, which describes the temporal evolution of the phase-shift $\Delta\phi$ between two areas when they are weakly interacting \cite{Kuramoto1984}:
\begin{equation}\label{eq:Gamma}
\frac{d(\Delta\phi)}{dt} = \Gamma(\Delta\phi)
\end{equation} 
The function $\Gamma(\Delta\phi)$ for the rate model is shown in the left panel of Fig.~\ref{fig:PRC}B. Stable phase lockings are given by zeroes of $\Gamma(\Delta\phi)$ with negative slope crossing and are surrounded by basins of attraction (i.e.~sets of configurations leading to a same equilibrium), whose boundaries are unstable in- and anti-phase lockings (Fig.~\ref{fig:PRC}A). For the network model, a function $\widetilde\Gamma(\Delta\phi)$ with an analogous interpretation and a similar shape, shown in the right panel of Fig.~\ref{fig:PRC}B, can be extracted from simulations, based on a phase description of ``LFP'' time-series (see {\it Methods} and Supporting~Figure~S1A). The analogous distribution of the zero-crossings of $\Gamma(\Delta\phi)$ and $\widetilde\Gamma(\Delta\phi)$ results in equivalent phase-locking behaviors for the rate and network models. Thus spontaneous symmetry breaking leads to multi-stability between alternative out-of-phase-lockings and to the emergence of unidirectional effective driving within a symmetric structural motif.

\subsection*{Control of directed causality} 
\begin{figure}[!ht]
\begin{center}
\includegraphics[width=12.35cm]{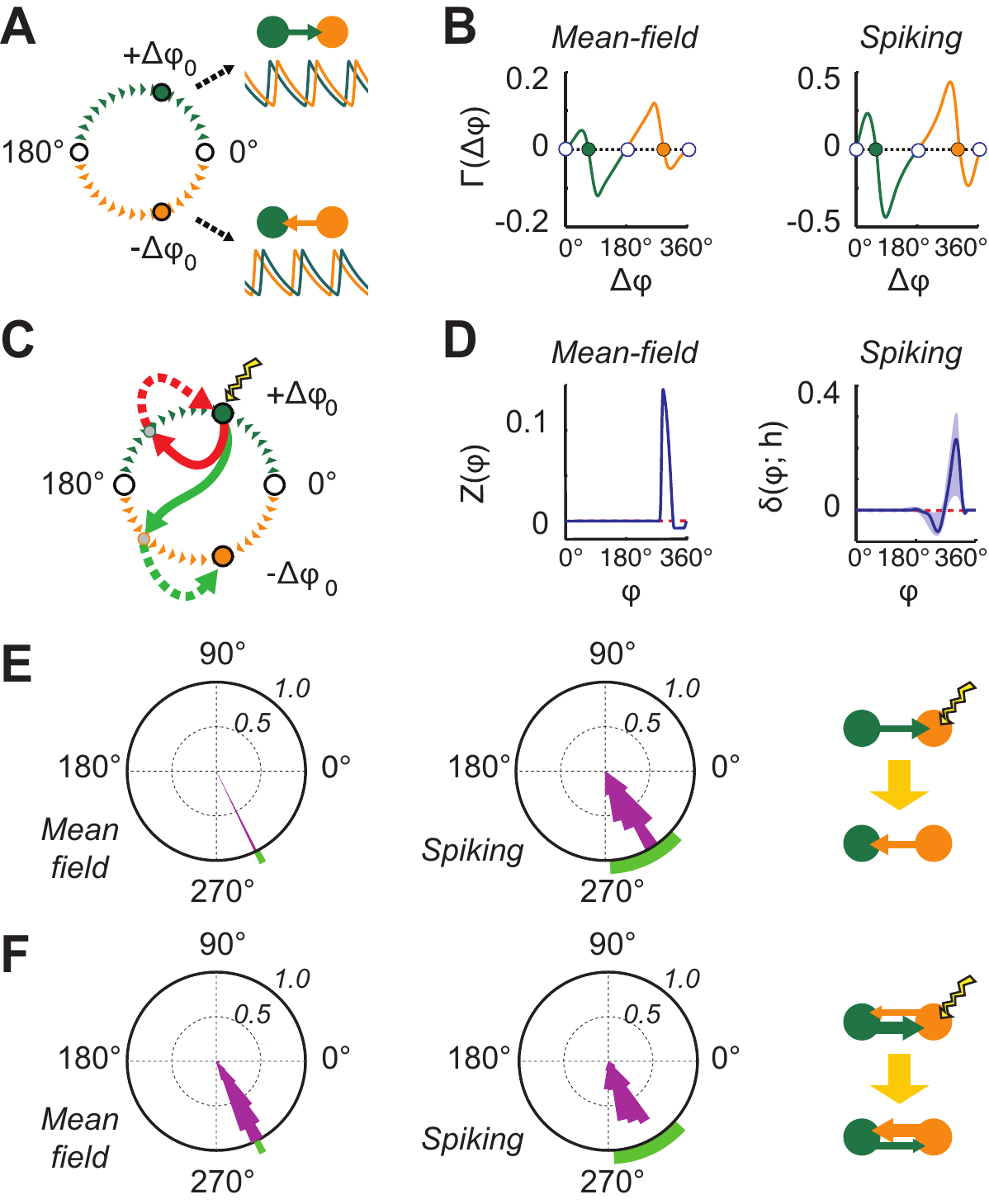} 
\end{center}
\caption{ }
\label{fig:PRC}
\end{figure}

Because of multi-stability, transitions between effective motifs within a family can be triggered by transient perturbations, without need for structural changes.  We theoretically determine conditions for such transitions to occur. The application of a pulse of current of small intensity $h$ advances or delays the phase $\phi$ of the ongoing local oscillation (see Supporting~Figure~S1B). This is true for rate oscillations of the mean-field rate model, but also for ``LFP'' oscillations reflecting rhythmic synchronization in the network model. In the latter case, the collective dynamics is perturbed by synchronously injecting pulse currents into all of the neurons within an area. The induced phase shift $\delta\phi$ depends on the perturbation strength $h$ but also on the phase $\varphi$ at which the perturbation is applied. For the network model, this $\delta\phi(\varphi; h)$ can be measured directly from numeric simulations of a perturbed dynamics (see \textit{Methods} and right panel of Fig.~\ref{fig:PRC}D). For the rate model, the phase shift induced by an instantaneous phased perturbation can be described analytically in terms of the Phase Response Curve (PRC) $Z(\phi) = \frac{\partial \phi}{\partial h}$ \cite{Kuramoto1984} (see Fig.~\ref{fig:PRC}D, left, and Supporting~Text~S1). After a pulse, the phase-shift between two areas is ``kicked out''  of the current equilibrium locking $\Delta\phi$ and assumes a new transient value $\Delta\phi^*$ (solid paths in Fig.~\ref{fig:PRC}C), which, for weak perturbations and inter-areal coupling, reads:
\begin{equation}\label{eq:PRC}
\Delta\phi^*=\Delta\phi + \delta\phi(\varphi; h) \mbox{\,\,\,[\,}\simeq \Delta\phi +hZ(\varphi)\mbox{\,]}
\end{equation}
where the approximate equality between square brackets holds for the mean-field rate model.
If $\Delta\phi^*$ falls into the basin of attraction of a different phase-locking configuration than $\Delta\phi$, the system will settle within few oscillation cycles into an effective connectivity motif with a different directionality (dashed green path in Fig.~\ref{fig:PRC}C). Even relatively small perturbations can induce an actual transition, if applied in selected narrow phase intervals in which the induced $\delta\phi(\phi; h)$ grows to large values. For most application phases, however, even relatively large perturbations fail to modify the effective driving direction (dashed red path in Fig.~\ref{fig:PRC}C), because the induced perturbation $\delta\phi(\phi; h)$ is vanishingly small over large phase intervals (Fig.~\ref{fig:PRC}D). This is a robust property, shared by the two (radically different) models we consider here and ---we hypothesize--- by any local circuit generating fast oscillations through a mechanism based on delayed mutual inhibition. 
As a consequence, for a given perturbation intensity, a successful switching to a different effective motif occurs only if the perturbation is applied within a specific phase interval, that can be determined analytically from the knowledge of $\Gamma(\Delta\phi)$ and of $Z(\phi)$ for the rate model, or semi-analytically from the knowledge of $\widetilde\Gamma(\Delta\phi)$ and $\delta\phi(\phi; h)$ (see {\it Methods}). Fig.~\ref{fig:PRC}E--F reports the fraction of simulated phased pulses that induced a change of effective directionality as a function of the phase of application of the perturbation. The phase intervals for successful switching predicted by the theory are highlighted in green. We performed simulations of the rate (Figs.~\ref{fig:PRC}E--F, left column) and of the network (Figs.~\ref{fig:PRC}E--F, middle column) models, for unidirectional (Figs.~\ref{fig:PRC}E) and leaky driving (Figs.~\ref{fig:PRC}F) effective motifs. Although our theory assumes small inter-areal coupling and is rigorous only for the rate model, the match between simulations and predictions is very good for both models and families of motifs. 

In Figs.~\ref{fig:PRC}E--F, we perturb the dynamics of the laggard area, but changes in directionality can also be achieved by perturbing the leader area (Supporting~Figure~S2). Note also that,  in the network model, direction switchings can take place spontaneously, due to noisy background inputs.  Such noise-induced transitions, however,  occur typically on time-scales of the order of seconds, i.e. slow in terms of biologic function, because the phase range for successful switching induction is narrow.

\subsection*{Effective entrainment} A second non-linear dynamic mechanism underlying the sequence of effective motifs of Figures~\ref{fig:EFFECTI_unidir}~and~\ref{fig:EFFECTI_leaky} is {\it effective entrainment}. In this phenomenon, the complex dynamics of neural activity seems intriguingly to be dictated by effective rather than by structural connectivity. 

\begin{figure}[!ht]
\begin{center}
\includegraphics[width=12.35cm]{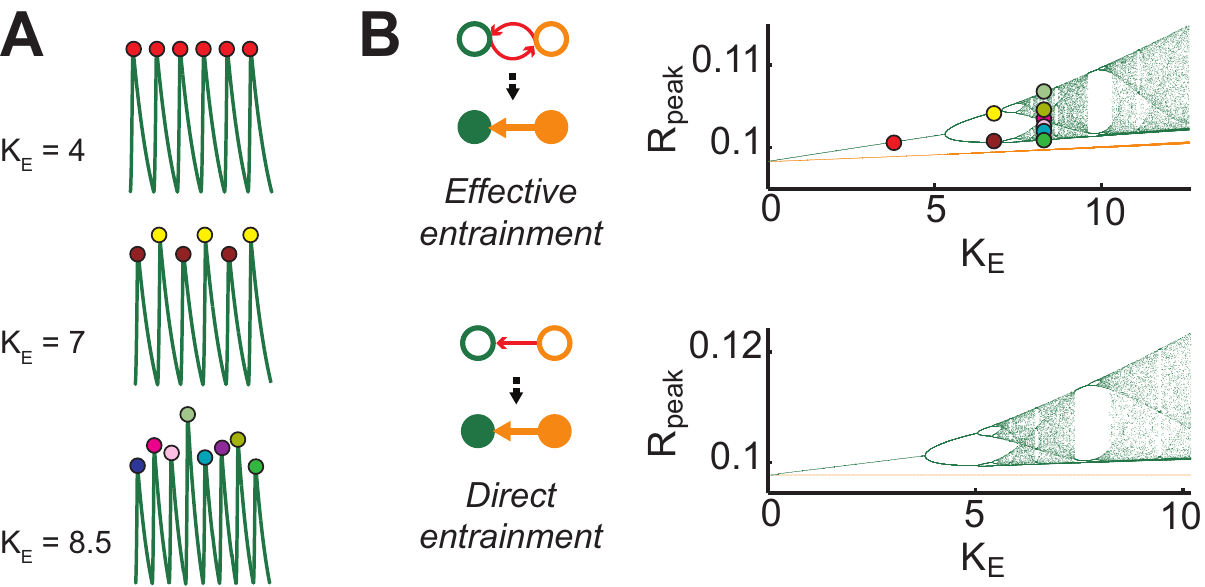} 
\end{center}
\caption{ }
\label{fig:BIFO}
\end{figure}

We consider as before a rate model of $N=2$ reciprocally connected areas (Fig.~\ref{fig:MODELS}D). In order to properly characterize effective entrainment, we review the concept of {\it bifurcation diagram} \cite{Schuster2005}. As shown in \cite{Battaglia2007}, when the inter-areal coupling $K_E$ is increased, rate oscillations become gradually more complex (cfr. Fig.~\ref{fig:BIFO}A), due to the onset of deterministic chaos (see also \cite{Battaglia2011} for a similar mechanism in a more complex network model). For small $K_E$, oscillations are simply periodic (e.g. $K_E = 4$). Then, for intermediate $K_E$ (e.g. $K_E = 7$),  the peak amplitudes of the laggard area oscillation assume in alternation a small number of possible values (period doubling). Finally, for larger $K_E$ (e.g. $K_E = 8.5$), the laggard peak amplitudes fluctuate in a random-like manner within a continuous range. This sequence of transitions can be visualized by plotting a dot for every observed value of the peak amplitudes of oscillation cycles, at different values of $K_E$. The accumulation of these dots traces an intricate branched structure, which constitutes the bifurcation diagram (Fig.~\ref{fig:BIFO}B). 

Bifurcation diagrams for the leader and for the laggard area are plotted in Fig.~\ref{fig:BIFO}B (top panel, in orange and green color, respectively). We compare these bifurcation diagrams with the analogous diagrams constructed in the case of two {\it unidirectionally} coupled oscillating areas. Qualitatively similar bifurcation sequences are associated to the dynamics of the laggard area (bidirectional coupling) and of the driven area (unidirectional coupling, Fig.~\ref{fig:BIFO}B, bottom panel, green color), \revs{for not too strong inter-areal couplings}. In the case of unidirectional coupling, the peak amplitudes of the unperturbed driver area oscillations do not fluctuate at all. Therefore, the corresponding bifurcation diagram is given by a constant line (Fig.~\ref{fig:BIFO}B, bottom panel, orange color). In the case of bidirectional coupling, the peak amplitudes of the leader area oscillations undergo fluctuations, but only with a tiny variance. Thus, the corresponding bifurcation diagram has still the appearance of a line, although now ``thick'' and curved (zooming would reveal bifurcating branches). Note that, for unidirectional coupling, the structural connectivity is explicitly asymmetric. The periodic forcing exerted by the driving area is then known to entrain the driven area into chaos \cite{Pikovsky2001}. Such direct entrainment is the {\it dynamical cause} of chaos. On the other hand, for bidirectional coupling, the structural connectivity is symmetric. However, due to spontaneous symmetry breaking, the resulting effective connectivity is asymmetric and the system behaves {\it as if} the leader area was a driver area, entraining the laggard area into chaos being only negligibly affected by its back-reaction. Such effective entrainment can be seen as an {\it emergent dynamical cause} of chaos. Thus, the dynamics of a symmetric structural motif with asymmetric effective connectivity and of a structural motif with a matching asymmetric topology are equivalent. 

\revs{For a sufficiently strong inter-areal coupling, symmetry in the dynamics of the bidirectional structural motif is suddenly restored \cite{Battaglia2007}, in correspondence with a transition to the mutual driving family of effective motifs (Figure~\ref{fig:EFFECTI_mutual}). As a result, in absence of symmetry breaking, effective driving cannot anymore take place. Thus, for a too strong inter-areal coupling, the emergent anisotropy of effective connectivity is lost, and, with it, the possibility of a dynamic control of effective connectivity (at least via the previously discussed strategies).}

\subsection*{Information follows causality} 
Despite its name, Transfer Entropy is not directly related to a transfer of information in the sense of neuronal information processing.
The TE from area $X$ to area $Y$ measures indeed just the degree to which the knowledge of the past ``LFP'' of $X$ reduces the uncertainty about the future ``LFP'' of $Y$ \cite{Kaiser2002, Barnett2009}. As a matter of fact, however, the information stored in neural representations must be encoded in terms of spikes, independently from the neural code used. Therefore, it is important to understand to which extent an effective connectivity analysis based on ``macroscopic'' dynamics (i.e. TEs estimated from ``LFPs'') can pretend to describe actual ``microscopic'' information transmission (i.e. at the level of spiking correlations).

\begin{figure}[!ht]
\begin{center}
\includegraphics[width=12.35cm]{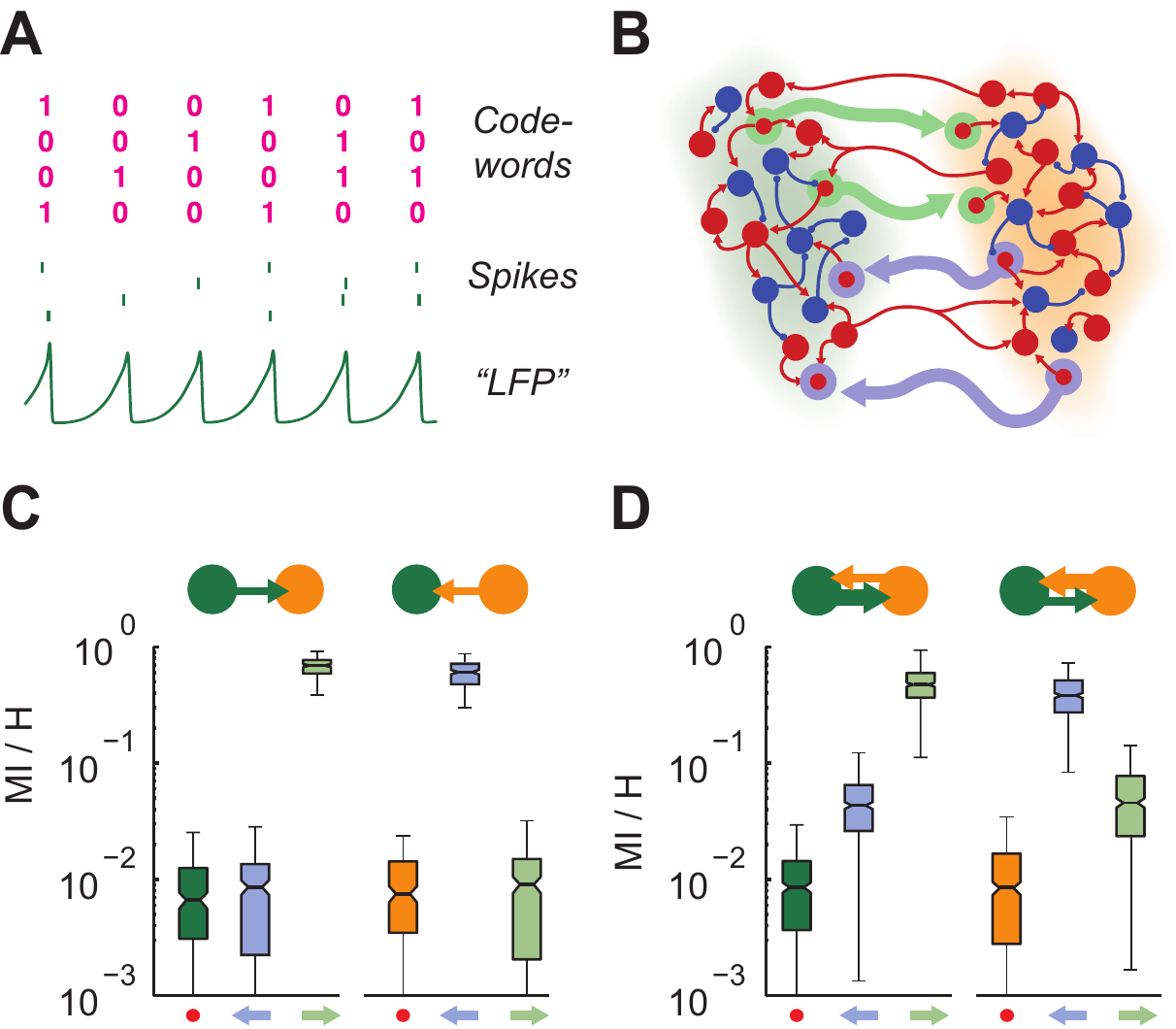} 
\end{center}
\caption{}
\label{fig:INFOTRANS}
\end{figure}

In order to address this issue, we first introduce a framework in which to quantify the amount of information exchanged by interacting areas. In the case of our model, {\it rate fluctuations} could encode only a limited amount of information, since firing rate oscillations are rather stereotyped. On the other hand, a larger amount of information could be encoded based on {\it spiking patterns}, since the spiking activity of single neurons is very irregular and thus characterized by a large entropy \cite{MacKay2003, Strong1998}. As illustrated by Fig.~\ref{fig:INFOTRANS}A, a code can be built, in which a ``1'' or a ``0'' symbol denote respectively firing or missed firing of a spike by a specific neuron at a given oscillation cycle. Based on such an encoding, the neural activity of a group of neurons is mapped to digital-like streams, ``clocked'' by the ongoing network rhythm, in which a different ``word'' is broadcast at each oscillation cycle. Note that we do not intend to claim that such a code is actually used in the brain. Nevertheless, we introduce it as a theoretical construct grounding a rigorous analysis of information transmission.

We focus here on the fully symmetric structural motif of $N=2$ areas of Fig.~\ref{fig:MODELS}C. We modify the network model considered in the previous sections by embedding into it {\it transmission lines} (TLs), i.e. mono-directional fiber tracts dedicated to inter-areal communication (see Fig.~\ref{fig:INFOTRANS}B). In more detail, selected sub-populations of {\it source} excitatory neurons within each area establish synaptic contacts with matching {\it target} excitatory or inhibitory cells in the other area, in a one-to-one cell arrangement. Synapses in a TL are strengthened with respect to usual synapses, \revs{by multiplying their peak conductance by a multiplier $K_{TL}$ (see {\it Methods} section). Such multiplier is selected to be large, but not too much, in order not to affect the phase-relations between the collective oscillations of the two areas. Indeed, selecting a too large $K_{TL}$ would lead to an in-phase-locking configuration in which collective dynamics is enslaved to the synchronous activity of source and target populations. As analyzed in the Supporting~Figure~S3, a suitable tuning of $K_{TL}$ ensures that source-to-target neuron communication is facilitated as much as possible, without disrupting the overall effective connectivity (associated to the unperturbed phase-locking pattern). Note that such TL synapses are here introduced as a heuristic device, allowing to maximize the potential capacity of inter-areal communication channels \cite{MacKay2003}. However, due to the occurrence of consistent spike-timing relations in out-of-phase locked populations, it might be that spike-timing-dependent plasticity \cite{Dan2004} lead to the gradual emergence of subsets of synapses with substantially enhanced weight \cite{Song2005}, which would play a role in inter-circuit communication very similar to TL synapses.} 

The information transmission efficiency of each TL, for the case of different effective motifs, is then assessed by quantifying the Mutual Information (MI) \cite{MacKay2003, Strong1998} between the ``digitized'' spike trains of pairs of source and target cells (see {\it Methods}). Since a source cell spikes on average every five or six oscillation cycles, the firing of a single neuron conveys $\mathrm{H} \simeq$ 0.7 bits of information per oscillation cycle. MI normalized by the source entropy H indicates how much of this information reaches the target cell, a normalized MI equal to unity denoting lossless transmission. As shown by Fig.~\ref{fig:INFOTRANS}C--D, the communication efficiency of embedded TLs depends strongly on the active effective motif. In the case of unidirectional driving effective motifs (Fig.~\ref{fig:INFOTRANS}C), communication is nearly optimal along the TL aligned with the effective connectivity. For the misaligned TL, however, no enhancement occurs with respect to control (i.e.~pairs of connected cells not belonging to a TL). In the case of leaky driving effective motifs (Fig.~\ref{fig:INFOTRANS}D), communication efficiency is boosted for both TLs, but more for the TL aligned with the dominant effective direction. For both families of effective motifs, despite the strong anisotropy, the communication efficiencies of the two embedded TLs can be ``swapped''  within one or two oscillation cycles, by reversing the active effective connectivity through a suitable transient perturbation (see Fig.~\ref{fig:PRC}E--F). The considered $N=2$ structural motif acts therefore as a ``diode'' through which information can propagate efficiently only in one (dynamically reconfigurable) direction determined by effective connectivity.

\section*{Discussion}

\subsection*{Mechanisms for effective connectivity switching}

We have shown that a simple structural motif of interacting brain areas can give rise to multiple effective motifs with different directionality and strengths of effective connectivity, \revs{organized into different families.} Such effective motifs correspond to distinct dynamical states of the underlying structural motif. Beyond this, dynamic multi-stability makes the controlled switching between effective motifs \revs{within a same family} possible without the need for any structural change. 

\revs{On the contrary, transitions between effective motifs belonging to different families (e.g. a transition from a unidirectional to a leaky driving motif) cannot take place without changes in the strength of the delay of inter-areal couplings, even if the overall topology of the underlying structural motif does not need to be modified. Each specific effective motif topology (i.e. motif family) is robust within broad ranges of synaptic conductances and latencies, however if parameters are set to be close to critical transition lines separating different dynamical regimes, transitions between different families might be triggered by moderate and unspecific parameter changes. This could be a potential role for neuromodulation, known to affect the net efficacy of excitatory transmission and whose effect on neural circuits can be modeled by coordinated changes in synaptic conductances \cite{Brunel2001, Seamans2004}.}

\revs{Note that} dynamical coordination of inter-areal interactions based on precisely-timed synchronous inputs would be compatible with experimental evidence of phase-coding \cite{OKeefe1993, Laurent1994, Lisman2005, Montemurro2008, Vinck2010, Nadasdy2010}, indicating a functional role for the timing of spikes relative to ongoing brain rhythms (stimulus-locked \cite{DeCharms1996, Arabzadeh2006} as well as stimulus-induced or spontaneous \cite{Koepsell2010}). Note also that the time of firing is potentially controllable with elevated precision \cite{Tiesinga2008, Tiesinga2010, Kayser2010} and has been found to depend on the phase of LFPs in local as well as in distant brain areas \cite{Canolty2010}. 

In general, control protocols different from the one proposed here might be implemented in the brain. For instance, phased pulses might be used as well to stabilize effective connectivity in the presence of stronger noise. Interestingly, the time periods framed by cycles of an ongoing oscillation can be sliced into distinct functional windows in which the application of the same perturbation produces different effects. 

\revs{Finally}, in addition to ``on demand'' transitions, \revs{triggered by exogenous ---sensory-driven--- or endogenous ---cognitive-driven--- control signals}, noise-driven switching between effective motifs might occur spontaneously, yielding complex patterns of activity during resting state \cite{Ghosh2008, Deco2009b, Deco2011}.

\subsection*{Transfer Entropy as a measure of effective connectivity}

As revealed by our discussion of spontaneous symmetry breaking and effective entrainment, an effective connectivity analysis based on TE provides a description of complex inter-areal interactions compliant with a dynamical systems perspective. It provides, thus, an intuitive representation of dynamical states that is in the same ``space'' as anatomical connectivity. \revs{Furthermore, as indicated by the analysis of Figure~\ref{fig:INFOTRANS}C-D, effective connectivity motifs obtained through TE point also at} the specific modalities of information routing enabled by \revs{the associated dynamical states.} In this sense, we can conclude that ``causality follows dynamics''. 

\revs{TE constitutes a model-free approach (although, non ``parameter-free'', cfr. forthcoming section and Figure~\ref{fig:BINLAG}) to the effective connectivity problem, suitable for exploratory data-driven analyses. In this sense it differs from regression-based methods like Granger Causality (GC) \cite{Granger1969, Ding2006} or from Dynamic Causal Modeling (DCM) \cite{Friston2003}, which are model-driven \cite{Friston2011, Bressler2011, ValdesSosa2011}. Strategies like DCM, in particular, assume prior knowledge about the inputs to the system and works by comparing the likelihood of different a priori hypotheses about interaction structures. Such an approach has the undeniable advantage of providing a description of actual mechanisms underlying effective connectivity changes (the stimulus-dependence of effective couplings is indeed modeled phenomenologically). However, it might be too restrictive (or arbitrary) when the required a-priori information is missing or highly uncertain. TE, on the contrary: does not require any hypothesis on the type of interaction;  should be able to detect even purely non-linear interactions and should be robust against linear cross-talk between signals \cite{Vicente2011}. These features, together with} the efficacy of TE for the causal analysis of synthetic time-series, advocate for a more widespread application of TE methods to real neural data \cite{Gourevitch2007, Besserve2010, Wibral2011} (\revs{at the moment limited by the need of very long time-series \cite{Vicente2011}).}

\revs{Note that we do not intend to claim superiority of TE in some general sense. As a matter of fact TE is equivalent to GC, as far as the statistics of the considered signals are gaussian \cite{Barnett2009}. Furthermore, non-linear generalizations of GC and DCM \cite{Gourevitch2006, Marinazzo2008, Stephan2008, Marinazzo2011} might be able to capture at a certain extent the complex self-organized dynamics of the neural activity models analyzed in the present study.  However, a systematic comparison of the performance of different methods in capturing causal connectivity of realistic non-linear models of neural dynamics goes beyond the focus of the present study and is deferred to future research.}

\revs{We finally would like to stress, to avoid any potential confusion, that the structural motifs analyzed in the present study are well distinct from causal graphical models of neural activity, in the statistical sense proper of DCMs \cite{Friston2003, Pearl2000}. They constitute indeed actual mechanistic models of interacting populations of spiking neurons, with a highly non-linear dynamics driven by background noise. Connections in these models are model synapses, i.e. mere structural couplings, not phenomenological effective couplings. Thus, effective connectivity is not constrained a priori, as in DCMs, but is an emergent property of network dynamics, consistent with the existence of effective motif topologies different from the underlying structural topology.}

\subsection*{Robustness of Transfer Entropy estimation}

\begin{figure}[!ht]
\begin{center}
\includegraphics[width=12.35cm]{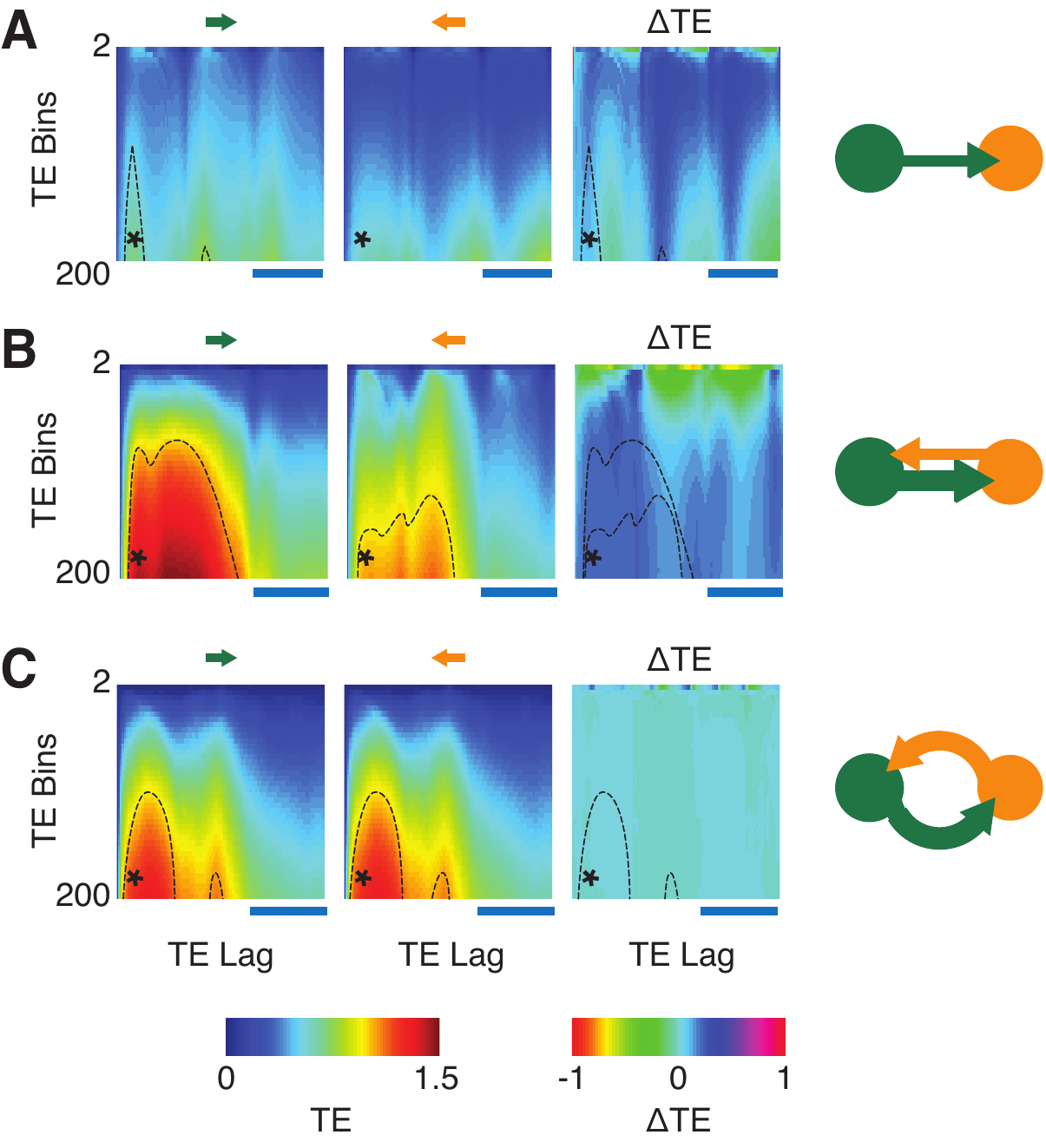} 
\end{center}
\caption{ }
\label{fig:BINLAG}
\end{figure}

\revs{The effective connectivity analyses presented in this study were conducted by evaluating TEs under specific parameter choices. However, absolute values of TE depend on parameters, like, notably, the resolution at which ``LFP'' signals are quantized and the time-lag at which we probe causal interactions. As discussed in detail in the {\it Methods} section, estimation of TE requires the sampling of joint distributions of ``LFP'' values in different areas at different times. Such distributions are sampled as histograms, based on discrete multi-dimensional binning. In practice, each ``LFP'' time-series is projected to a stream of symbols from a discrete alphabet, corresponding to different quantization levels of the continuous ``LFP'' signals \cite{Staniek2008}. The actual number $B$ of used bins is a free parameter, although some guiding criteria for its selection do exist \cite{Kaiser2002}. Concerning time-lag $\tau$, our TE analysis (conducted at the first Markov order \cite{Schreiber2000}, following \cite{Honey2007, Besserve2010}) describes predictability of ``LFPs'' at time $t$ based on ``LFPs'' at time $t-\tau$. The used time-lag $\tau$ is once again a free parameter. To deal with this arbitrariness in parameter choices, we explore systematically the dependence of TE estimations from the aforementioned parameters, by varying both $B$ and $\tau$ in a wide continuous range. Figure~\ref{fig:BINLAG} summarizes the results of this analysis, for three different effective motifs.} 

\revs{Considering the dependency on time-lag $\tau$, a periodic structure is clearly noticeable in the TE matrices reported in Figure~\ref{fig:BINLAG}. TE values tend to peak in precise bands of $\tau$, related to latencies between the oscillations of different areas.  The analysis of the unidirectional driving motif (Figure~\ref{fig:BINLAG}A), associated to leader-laggard periodic configurations, is particularly transparent (and has a high pedagogic value). Two characteristic time-lags can be defined: a ``short'' lag $\tau_{XY}$, given by the time-shift from oscillation peaks of the leader area $X$ to oscillation peaks of the laggard area $Y$; and a ``long'' lag, $\tau_{YX} = T -\tau_{XY}$, given by the time-shift from laggard to leader oscillation peaks (here, $T$ is an average oscillation period, common to both areas leader and laggard areas $X$ and $Y$). TE in the direction from leader to laggard, $TE_{XY}$, peaks for the first time at a time-lag $\tau=\tau_{XY}$ (and then at lags $\tau_{XY}+nT$, where $n$ is a positive integer). TE in the direction from laggard to leader, $TE_{YX}$, peaks first at a time-lag $\tau=\tau_{YX}$ (and then at lags $\tau_{YX}+nT$). If the ``LFP'' signals were deterministic and {\it strictly} periodic, the quantities $TE_{XY}(\tau_{XY})$ and $TE_{YX}(\tau_{YX})$ would be identical (and diverging for infinite precision \cite{Schreiber2000}). However ``LFP'' signals are only periodic {\it on average} and have a stochastic component, due to the joint effect of random network connectivity and noisy background inputs. This stochastic component is responsible for small cycle-to-cycle fluctuations in the amplitude of ``LFP'' oscillation peaks. As discussed more in depth in a next subsection, the efficiency with which fluctuations in the output of a local area can induce (i.e., can ``cause'') fluctuations of the output of a distant interconnected area depends on the instantaneous local excitability of this target area, which is undergoing a rhythmic modulation due to the ongoing collective oscillation \cite{Fries2005, Volgushev1998}. As a result,  TE can reach different peak values in different directions (and, as a matter of fact, $TE_{XY}(\tau_{XY}) > TE_{YX}(\tau_{YX})$).}

\revs{Considering then the dependence on signal quantization, we observe that TE values tend to grow for increasing number of bins $B$, i.e. for a finer resolution in tracking ``LFP'' amplitude variations. This can be once again understood in terms of the temporal structure of ``LFP'' signals. As just mentioned, dynamic correlations between small ``LFP'' amplitude fluctuations carry information relevant for causality estimation. This information would be completely lost by using a too small number of bins for TE evaluation, given that the largest contribution to the dynamic range of ``LFP'' signals is provided by its fairly stereotyped oscillatory component. Conversely, using a too large number of bins would lead to under-sampling artifacts (therefore, we do not consider the use of more than $B=200$ quantization bins). }

\revs{By evaluating a threshold for statistical significance independently for each direction and combination of $B$ and $\tau$, we find that, for weak inter-areal coupling, TE never goes above this threshold in the laggard-to-leader direction  (Figure~\ref{fig:BINLAG}A). We are also unable to find any choice of $B$ and $\tau$ such that, for intermediate inter-areal coupling,  TE in the laggard-to-leader direction becomes larger or equal than TE in leader-to-laggard direction (Figure~\ref{fig:BINLAG}B). Looking at matrices of the causal unbalancing $\Delta\mathrm{TE}$ (see {\it Methods}, and Figure~\ref{fig:BINLAG}, third column), we see indeed that, for weak and intermediate coupling strengths, effective connectivity is {\it robustly asymmetric} in the parameter regions in which causal interactions are statistically significant. Effective connectivity is on the contrary balanced for strong inter-areal coupling (Figure~\ref{fig:BINLAG}C).}

\revs{We can thus summarize the previous statements by saying that absolute values of TE depend on the choices of $B$ and $\tau$, but that the topology of the resulting effective motif does not (at least in the wide range considered for this robustness analysis).}

\subsection*{Self-organized control of communication-through-coherence}

Traditionally, studies about \revs{communication-through-coherence or} long-range binding between distant cell assemblies have emphasized the importance of in-phase locking (see, e.g. \cite{Womelsdorf2007, Roelfsema1997}). Although, as previously mentioned, in-phase locking (as well as anti-phase locking) can also arise in our models for different choices of coupling delays and inhibition strengths \cite{Battaglia2007}, we decided in the present study to focus on out-of-phase lockings. The case of spontaneous symmetry breaking is indeed particularly interesting, because it underlie the emergence of a dominant directionality in the causal influences between areas reciprocally coupled with comparable strengths. \revs{Furthermore, spontaneous symmetry breaking is responsible for the multi-stability between effective connectivity configurations, thus opening the way to a self-organized control of inter-areal interactions \cite{Bressler2001, Varela2001}.}

 \revs{In particular,} our study confirms that the reorganization of oscillatory coherence might regulate the relative weight of bottom-up and top-down inter-areal influences \cite{Gilbert2007, Engel2001} or select different interaction modes within cortical networks involving areas of similar hierarchical level, as in the case of motor preparation or planning \cite{Brovelli2004, Westendorff2010} or language \cite{Mainy2008}.
 
As a next step, we directly verified that ``information follows causality'', since changes in effective connectivity \revs{are paralleled by reconfiguration of inter-areal communication modalities}. Following \cite{Fries2007, Womelsdorf2007}, we explain the anisotropic modulations of communication efficiency (see Fig.~\ref{fig:INFOTRANS}) in terms of a {\it communication-through-coherence} mechanism. In fact, because of the out-of-phase locking between rhythms, spikes emitted by neurons in a phase-leading area reach neurons in a phase-lagging area at a favorable phase in which they are highly excitable. Conversely, spikes emitted by neurons in a phase-lagging area reach neurons in a phase-leading area when they are strongly hyperpolarized by a preceding peak of synchronous inhibition.
\revs{This same mechanism underlie also the anisotropy of ``LFP''-based TE, since ``LFP'' fluctuations are the manifestation (at least in our model) of local population firing rate fluctuations.}

\revs{Therefore, by combining TE analyses of  ``LFP''-based effective connectivity with MI analyses of spike-based information transmission, we are able to establish a tight link between control of effective connectivity and control of communication-through-coherence, both of them being emergent manifestations of the self-organized dynamics of interacting brain rhythms.}

 \revs{To conclude, we also note that} similar mechanisms might be used beyond the mesoscale level addressed here. Multi-stabilities of structural motifs might be preserved when such motifs are interlaced as modules of a network at the whole-brain level \cite{Sporns2004}. Likewise, dynamic control of \revs{information routing} between neuronal clusters \cite{Song2005, Perin2011} or even single cells might occur within more local microcircuits \cite{Aertsen1989, Kispersky2011}.

\subsection*{Communication-through-coherence beyond rate coding}

\revs{The previous discussions suggest that} oscillations, \revs{rather than playing} a direct role in the representation of information, would be instrumental to the reconfigurable routing of information encoded in \revs{spiking activity. Original formulations of the communication-through-coherence hypothesis \cite{Fries2005} suggested that oscillatory coherence facilitates the transmission of local fluctuations of firing rate to a distant site, thus assuming implicitly a rate-based encoding of information in neuronal activity. However, more complex coding mechanisms based on patterns of precisely timed spikes might be achievable by biologically-plausible neuronal circuits \cite{Tiesinga2008, Tiesinga2010}.}

As a matter of fact, our study \revs{reveals} that the inherent advantages of ``labelled-line'' codes \cite{Reich2001, Rolls2011} (in which the information about which local neuron is firing is preserved) ---i.e., notably, an augmented information capacity with respect to ``summed-population'' codes--- might be combined with the flexibility and the reliability offered by the communication-through-coherence framework. \revs{Indeed, as shown by the analyses of Figure~\ref{fig:INFOTRANS}, suitable inter-areal phase relations make possible the transmission of information encoded in detailed spiking correlations, rather than just in population firing rate fluctuations.}

This is particularly interesting, since many cortical rhythms are only sparsely synchronized, with synchronous oscillations evident in LFP, Multi-Unit Activity or intracellular recordings but not in single unit spike trains \cite{Jarvis2001, Muresan2008, Yu2010}. \revs{Such sparse firing might possibly reflect population-coding of behaviorally-relevant information transcending rate-based representations \cite{MacKay1952, Mainen1995, Reich2001, Osborne2008, Ohiorhenuan2010}. Independently from the complexity of these hypothetic representations, our study shows that self-organized communication-through-coherence would have the sufficient potential to dynamically route the rich information that these representations might convey.}

\subsection*{Perspectives}

It is very plausible that flexible inter-areal coordination is achieved in the brain through dynamic self-organization \cite{Bressler2001} as in our models. However, qualitatively different mechanisms than symmetry breaking might contribute to the generation of dynamic effective connectivity in other regimes of activity. Despite sparse synchronization, the level of coherence in our model neuronal activity is larger than in many brain oscillations. However, our results might be generalized to activity regimes in which synchronization is weaker. Phase-relations have been shown to impact effective connectivity even in essentially asynchronous regimes \cite{Buehlmann2010}. It would be interesting to understand whether the dominant directionality of effective connectivity can be controlled when out-of-phase locking is only transient \cite{Honey2007, Varela2001}. 

Another open question is whether our theory can be extended to encompass the control of effective connectivity across multiple frequency bands \cite{Besserve2010}. This is an important question since top-down and bottom-up inter-areal communication might exploit different frequency channels, possibly due to different anatomic origins of ascending and descending cortico-cortical connections \cite{Buffalo2011}. 

Finally, we are confident that our theory might inspire novel experiments, attempting to manipulate the directionality of inter-areal influences via local stimulation applied conditionally to the phase of ongoing brain rhythms. \revs{Precisely timed perturbing inputs could indeed potentially be applied using techniques like electric \cite{Histed2009} or optogenetic \cite{Yizhar2011} microstimulation, especially in closed-loop implementations with millisecond precision \cite{Venkatraman2009, Leifer2011}. }

\section*{Methods}

\subsection*{Network model} 
Each area is represented by a random network of $n_E=4000$ excitatory and $n_I=4000$ inhibitory Wang-Buzs\'aki-type conductance-based neurons \cite{Wang1996}. The Wang-Buzs\'aki model is described by a single compartment endowed with sodium and potassium currents. \revs{Note that results (not shown) of simulations performed with a more realistic ratio of $n_E=4000$ excitatory and $n_I=1000$ inhibitory neurons per population would lead to qualitatively similar results with small parameter adjustments (using, for instance, parameters as in \cite{Battaglia2011}).} 

The membrane potential is given by:
\begin{equation}
C\frac{dV}{dt} =-I_L -I_{Na} -I_K +I_{ext} +I_{rec}
\end{equation}
where $C$ is the capacitance of the neuron, $I_L$ is a leakage current, $I_{ext}$ is an external noisy driving current \revs{(due to background Poisson synaptic bombardment)}, and $I_{Na}$ and $I_{K}$ are respectively a sodium and a potassium current, depending non linearly on voltage. The last input term $I_{rec}$ is due to recurrent interactions with other neurons in the network. Excitatory synapses are of the AMPA-type and inhibitory synapses of the GABA$_{\mbox{{\tiny A}}}$-type and are modeled as time-dependent conductances. A complete description of the model and a list of all its parameters are given in the Supporting~Text~S1.  ``LFP'' $\Lambda(t)=\langle V(t) \rangle$ is defined as the average membrane potential over the $N_E+N_I$ cells in each area. 

Short-range connections within a local area $k$ from population $\alpha_k$ to population $\beta_k$ are established randomly with
probability $p^{kk}_{\alpha, \beta}$ , where $\alpha$ and $\beta$ can be either one of the type $E$ (excitatory) or $I$. 
The excitatory populations $E_k$are allowed also to establish connections toward populations $E_l$ and $I_l$ in remote areas ($k \neq l$).
Such long-range connections are established with a probability $p_{E\alpha}^{kl}$ ($\alpha = E, I$). For simplicity, however, we
assume that $p_{II}^{kk} = p_{IE}^{kk} = p_I$ and that $p_{EE}^{kk} = p_{EI}^{kk} = p_{EE}^{kl} = p_{EI}^{kl} = p_E$. For each of the considered dynamical states, probabilities of connection are provided in the corresponding figure caption.

\subsection*{Network model with embedded transmission lines (TLs)}
First, a structural motif of interconnected random networks of spiking neurons is generated, as in the previous section. Then, on top of the existing excitatory long-range connections, additional stronger long-range connections are introduced in order to form directed transmission lines.
In each area a source sub-population, made out of 400 excitatory neurons, and a non-overlapping target sub-population, made out of 200 excitatory and 200 inhibitory neurons, are selected
randomly. Excitatory cells in the source populations get connected to cells in the target sub-populations of the other area via strong synapses. These connections are established in a one-to-one arrangement (e.g. each source cell establishes a TL-synapse with a single target cell that does not receive on its turn any other TL-synapse). 

The peak conductance $g_{TL}$ of TL-synapses is  $K_{TL}$ times stronger than the normal excitatory peak conductance $g_{E}$. For the simulations with TL (Fig.~\ref{fig:INFOTRANS} of the main paper), we set $K_{TL}$ = 22 and 24.5 respectively for the unidirectional and for the leaky driving effective motifs. Such unrealistically strong peak conductances, whose purpose is to optimize information transfer by enhancing spiking correlations, can be justified by supposing that each source neuron establishes multiple weaker synaptic contacts with the same target neuron. \revs{The multiplier $K_{TL}$ is selected to be as large as possible without altering the original out-of-phase locking relations between the two populations (Figure~S3A). Concretely, $K_{TL}$ is tuned by raising it gradually until when a critical point is reached in which the populations lock in-phase (Figure~S3C). Then, $K_{TL}$ is set to be just below this critical point (Figure~S3B).}

\subsection*{Rate model} 
Each area is represented by a single rate unit. The dynamical equations for the evolution of the average firing rate $R_k(t)$ in an area $k$ are given by:
\begin{eqnarray}\label{eq:ratemodel}
\dot{R_k}(t) = -R_k(t) + [I + K_I R_k(t-D)+ \\\nonumber
+ \sum_{l\neq k} K_E R_l(t-\bar{D})]_+,\,\,\,\,k,l = 1\ldots N
\end{eqnarray} 
Here, $[x]_+ = x$ if $X\geq 0$, and zero otherwise. A constant current $I$ represents a background input, $K_I$ stands for the strength of intra-areal inhibition, $K_E$ for the strength of inter-areal excitation and $D$ and $\overline{D}$ are the delays of the local and long-range interactions, respectively. We consider in this study only fully symmetric structural motifs of $N$ mutually connected areas. For each of the considered dynamical states, the values of  $K_I $, $K_E$, $D$ and $\overline{D}$ are provided in the figure caption.

\subsection*{Phase reduction and response} 
Given an oscillatory time-series of neuronal activity, generated indifferently by a rate or by a network model, a phase $\phi(t) =360^\circ\cdot(t-t_{max, \ell})/(t_{max,\ell+1}-t_{max,\ell})$, for $t_{max,\ell} \leq t < t_{max,\ell+1}$, is linearly interpolated over each oscillation cycle. Here  $t_{max,\ell}$ denotes the start time of the $\ell$-th oscillation cycle. Note that this definition does not require that the oscillation is periodic: this empiric phase ``elastically'' adapts to fluctuations in the duration of oscillation cycles (see Supporting~Figure~S1A).  

The phase shift induced by a pulse perturbation $I = h\delta(\phi-\phi_0)$ (see Supporting~Figure~S1B) is described by the Phase Response Curve (PRC) $Z(\phi) = \partial\phi/\partial h$ (see Eq. \eqref{eq:PRC} and \cite{Kuramoto1984}). For the rate model, the PRC can be evaluated analytically if certain general conditions on the relation between the oscillation period $T$ and the local inhibition delay $D$ are fulfilled \cite{Battaglia2007}. Analytical expressions for the PRC of the rate model, as plotted in Fig.~\ref{fig:PRC}D (left), are reported in the Supporting~Text~S1.

In the network model, it is possible to evaluate the phase-shift induced by a perturbation, by directly simulating the effects of this perturbation on the oscillatory dynamics. A perturbation consists of a pulse current of strength $h$ injected synchronously into all neurons of one area at a phase $\phi$ of the ongoing local oscillation. The induced phase-shift $\delta\phi(\phi; h)$ is estimated by comparing the phases of the perturbed and of the unperturbed oscillations, when a new equilibrium is reached after the application of the perturbation. In detail, since the ``LFP''  time-series are not strictly periodic and the phase relation is fixed only on average, the average time-lag between the perturbed and the unperturbed ``LFPs'' is measured by computing their crosscorrelogram over 50 oscillation cycles, starting from the 10-th cycle after the perturbation. This average time lag (readable from the position of the crosscorrelogram peak) is then translated into a phase-shift, by dividing it by the average period (estimated through autocorrelation analysis of the perturbed and unperturbed time-series over the same observation window). Vanishingly small perturbations do not induce long-lasting phase-shifts. Therefore, moderately large perturbation strengths have to be used. In this case, the dependence of $\delta\phi$ on $h$ is sensibly non-linear. 
As a consequence, we evaluate directly the resulting $\delta\phi(\phi; h)$ for the used perturbation strength $h$, plotted in Fig.~\ref{fig:PRC}D (right). The qualitative shape of $\delta\phi(\phi; h)$ however does not depend strongly on $h$. In particular, changes of $h$ affect the amplitude of the maximum phase-shift but not the perturbation phase for which it occurs. The curve $\delta\phi(\phi; h)$ is evaluated point-wise by applying perturbations at 100 different phases within a cycle. For each given phase, the perturbation is applied 100 times to 100 different cycles and the
corresponding phase-shifts are averaged. Confidence intervals for $\delta\phi(\phi; h)$ are determined phase-by-phase by finding the 2.5-th and the 97.5-th percentile of the induced phase-shift distribution across these 100 trials. 

\subsection*{Phase locking} 
For simplicity, we focus in the following on the case of $N=2$ areas, although our approach can be extended to larger motifs. The instantaneous phase-difference between two areas $X$ and $Y$ is given by $\Delta\phi(t) = \mod [\phi_X(t) - \phi_Y(t), 360^\circ]$. For vanishing inter-areal coupling, the time evolution of $\Delta\phi(t)$ is described by Eq. \eqref{eq:Gamma}. The term $\Gamma(\Delta\phi)$ is a functional of the phase response and of the limit cycle waveform of the uncoupled oscillating areas. For the rate model, $\Gamma(\Delta\phi)$ is determined from analytic expressions of $Z(\phi)$ and of the rate oscillation limit cycle $R(\phi)$ (note that the dependence on $t$ is replaced by a dependence on $\phi$ after phase-reduction) for $K_E = 0$. It can be expressed as $\Gamma(\Delta\phi) = C(\Delta\phi) - C(-\Delta\phi)$, with:
\begin{equation}
C(\Delta\phi) =  \int_{0^\circ}^{360^\circ} Z(\phi) R(\phi + \Delta\phi - D)d\phi
\end{equation}
The resulting expression is reported in the Supporting~Text~S1 and plotted in  Fig.~\ref{fig:PRC}B (left). Given Eq. \eqref{eq:Gamma}, the phase shifts $\pm\Delta\phi_0$ between the two areas $X$ and $Y$ in stable phase-locked states correspond to top-down zero-crossings of the functional $\Gamma(\Delta\phi)$ (i.e. zeroes with negative tangent slope, $\Gamma' < 0$). 

For the network model, the waveform of ``LFP'' oscillations can be determined through simulations. Since not all oscillation cycles are identical, the limit cycle waveform is averaged over 100 different cycles ---as for the determination of $\delta\phi(\phi; h)$--- to yield an average limit cycle $\langle \Lambda(\phi)\rangle$. Then, it is possible to evaluate a functional $\tilde{\Gamma}(\Delta\phi; h) = \tilde{C}(\Delta\phi; h) - \tilde{C}(-\Delta\phi; h)$, where:
\begin{equation}
\tilde{C}(\Delta\phi) =  \int_{0^\circ}^{360^\circ} \delta(\phi; h) \langle\Lambda(\phi + \Delta\phi - D)\rangle d\phi
\end{equation}
The functional $\tilde{\Gamma}(\Delta\phi; h)$ is plotted in Fig.~\ref{fig:PRC}B (right) for the used perturbation strength $h$. Although Eq. \eqref{eq:Gamma} does not exactly hold for the network model, the top-down zero-crossings of the functional $\tilde{\Gamma}(\Delta\phi; h)$ (whose position only weakly depends on $h$) continue to provide an approximation of the phase shifts $\pm \Delta\phi_0$ between the two areas $X$ and $Y$ in stable phase-locked states. In particular it is possible to predict whether the stable lockings will be in-phase, anti-phase or out-of-phase.

\subsection*{Phase intervals for effective connectivity switching}
Phase intervals in which the application of a pulsed perturbation leads to a change of effective connectivity directionality are determined theoretically as shown below. For $N=2$ and in a given phase-locking state, the phase of the leader area can be written as $\phi_{leader} = \varphi + \Delta\phi_0$ and the phase of the laggard area as $\phi_{laggard} = \varphi$. The application of a pulse perturbation of strength $h$ to the laggard area  shifts the phase of the ongoing local oscillation to $\phi_{laggard}^* \simeq \varphi + \delta(\varphi; h)$, where $\delta(\varphi; h)\simeq hZ(\varphi)$ holds for the rate model in the case of small perturbations. If the achieved transient phase-shift between the two areas, $\Delta\phi^* \simeq \Delta\phi_0 - \delta(\varphi; h)$, is falling into the basin of attraction of an alternative stable phase-locking (see Fig.~\ref{fig:PRC}C), then a switching toward a different effective motif takes place. Considering the dynamics of the instantaneous phase-shift, determined by the functionals $\Gamma(\Delta\phi)$ for the rate model and $\tilde{\Gamma}(\Delta\phi; h)$ for the network model (see Fig.~\ref{fig:PRC}B), switching will occur when:
\begin{equation}\label{eq:switching}
\Delta\phi_0 < \delta(\varphi; h) < \Delta\phi_0 + 180^\circ
\end{equation}
Here, we consider perturbations which induce a phase advancement, because the positive part of both the PRC in the rate model and the empiric 
$\delta\phi(\phi; h)$ in the network model is larger than the negative part (see Fig.~\ref{fig:PRC}D). 
For a fixed perturbation intensity $h$, the condition \eqref{eq:switching} will be fulfilled only if when the phase $\varphi$ of application of the perturbation falls within specific intervals, determined by the precise form of $\delta(\varphi; h)$.  These intervals are highlighted in green in Fig.~\ref{fig:PRC}E and F. Analogous considerations can be done in order to determine the intervals for successful switching when perturbing the leader area (see Supporting~Figure~S2).

\subsection*{Transfer Entropy (TE)} 
Let us consider first a structural motif with $N=2$ areas. Let $\Lambda_{X}(t)$ and $\Lambda_{Y}(t)$ be the ``LFP'' time-series of the two areas $X$ and $Y$, and let quantize them into $B$ discrete levels $\ell_1, \ldots, \ell_B$ (bins are equally sized). The continuous-valued ``LFP'' time-series are thus converted into strings of symbols $\tilde\Lambda_{X}(t)$ and $\tilde\Lambda_{Y}(t)$ from a small alphabet \cite{Staniek2008}.
Two transition probability matrices, $\left(P_{XY, Y}(\tau)\right)_{ijk} = P[\tilde\Lambda_Y(t+\tau) = \ell_i | \tilde\Lambda_Y(t) = \ell_j, \tilde\Lambda_X(t) = \ell_k]$ and $\left(P_{Y, Y}(\tau)\right)_{ij} = P[\tilde\Lambda_Y(t+\tau) = \ell_i | \tilde\Lambda_Y(t) = \ell_j]$, where the lag $\tau$ is an arbitrary temporal scale on which causal interactions are probed, are then sampled as normalized multi-dimensional histograms over very long symbolic sequences. These probabilities are sampled separately for each specific fixed phase-locking configuration. Epochs in which the system switches to a different phase-locking configuration, as well as transients following state switchings are dropped. The evaluation of $P_{XY, Y}(\tau)$ and $P_{Y, Y}(\tau)$ is thus based on disconnected symbolic subsequences, including overall $O(10^4)$ oscillation cycles. Then, following \cite{Schreiber2000}, the causal influence $\mathrm{TE}_{XY} (\tau)$ of area $X$ on area $Y$ is defined as the Transfer Entropy:
\begin{equation} \label{eq:TE}
\mathrm{TE}_{XY} (\tau) = \sum P_{XY, Y}(\tau) \log_2\frac{P_{XY,Y}(\tau)}{P_{Y,Y}(\tau)}
\end{equation}
where the sum runs over all the three indices $i$, $j$ and $k$ of the transition matrices. 

This quantity represents the Kullback-Leibler divergence \cite{MacKay2003} between the transition matrices $P_{XY, Y}(\tau)$ and $P_{Y, Y}(\tau)$, analogous to a distance between probability distributions. Therefore, $\mathrm{TE}_{XY} (\tau)$ will vanish if and only if $P_{XY, Y}(\tau)$ and $P_{Y, Y}(\tau)$ coincide, i.e. if the transition probabilities between different ``LFP'' values of area $Y$ do not  depend on past ``LFP'' values of area $X$. Conversely, this quantity will be strictly positive if these two transition matrices differ, i.e. if the past ``LFP'' values of area $X$ affect the evolution of the ``LFP''  in area $Y$.

\revs{We also measure the causal unbalancing \cite{Gourevitch2007}:}
\begin{equation}
\Delta\mathrm{TE} = \frac{\mathrm{TE}_{XY} - \mathrm{TE}_{YX}}{\mathrm{TE}_{XY}+\mathrm{TE}_{YX}}
\end{equation}
\revs{which is normalized in the range $-1 \le \Delta\mathrm{TE} \le 1$. A value close to zero denotes symmetric causal influences in the two directions, while large absolute values of $\Delta\mathrm{TE}$ signal the emergence of asymmetric effective connectivity motifs.}

\subsection*{Partialized Transfer Entropy (pTE)} 
Considering now a structural motif with $N=3$ areas, equation \eqref{eq:TE} has to be modified in order to distinguish causal interactions which are direct (e.g. $X$ toward $Y$) from interactions which are indirect (e.g. $X$ toward $Y$, but through $Z$). A solution allowing to assess only direct causal influences is partialization \cite{Schreiber2000, Barnett2009}. Indirect interactions from area $X$ to area $Y$ involving a third intermediate area $Z$ are filtered out by conditioning the transition matrices for the ``LFP'' activity of $Y$ with resepect to the activity of the $Z$. Two conditional transition matrices,  $\left(P_{XY, Y | Z}(\tau)\right)_{ijkl} = P[\tilde\Lambda_Y(t+\tau) = \ell_i | \tilde\Lambda_Y(t) = \ell_j, \tilde\Lambda_X(t) = \ell_k, \tilde\Lambda_Z(t) = \ell_l]$ and $\left(P_{Y, Y | Z}(\tau)\right)_{ijl} = P[\tilde\Lambda_Y(t+\tau) = \ell_i | \tilde\Lambda_Y(t) = \ell_j, \tilde\Lambda_Z(t) = \ell_l]$, are then constructed and used to evaluate:
\begin{equation} \label{eq:TE3}
\mathrm{TE}_{XY | Z} (\tau) = \sum P_{XY, Y | Z}(\tau) \log_2\frac{P_{XY,Y | Z}(\tau)}{P_{Y,Y | Z}(\tau)}
\end{equation}
where the sum runs over all the four indices $i$, $j$, $k$ and $l$. The effective connectivity in the panels C of Figures~\ref{fig:EFFECTI_unidir},~\ref{fig:EFFECTI_leaky}~and~\ref{fig:EFFECTI_mutual} is computed using pTE according to equation \eqref{eq:TE3}.

\subsection*{Statistic validation of effective connectivity} Absolute values of $\mathrm{TE}_{XY}$ depend strongly on the time-lag $\tau>0$ and on the number of discrete levels $B$. Nevertheless, we find that relative strengths of causal influences are qualitatively unchanged over broad ranges of parameters, as displayed in the Supporting~Figure~S1. 
Furthermore the ``plug-in'' estimates of TE given by equations \eqref{eq:TE} and \eqref{eq:TE3} suffer from finite-sampling biases, and a rigorous debiasing procedure is not yet known \cite{Kaiser2002}. Therefore, for each value of $\tau$ and $B$ it is necessary to assess the significancy of the inferred causal interactions through comparison with suitably randomly resampled data \cite{Efron1993}. To estimate the confidence intervals for the estimated TEs and the baseline for significancy we adopt a geometric bootstrap method \cite{Politis1994}, guaranteed to generate resampled time-series with similar auto- and cross-correlation properties up to a certain lag. This is important, since ``LFP'' time-series have a strong oscillatory component, whose correlation structure has to be maintained under resampling. Each resampled time-series $\Lambda^{bs}_X(t)$ consists of a concatenation of blocks sampled from the original time-series $\Lambda_X(t)$. Each $\Lambda^{bs}_X(t)$ has the same length as the original $\Lambda_X(t)$.
Every upward crossing, i.e. every time at which $\Lambda_X(t)$ crosses from below its time-averaged value $\overline{\Lambda_X(t)}$, is a potential start-time for a block. The first element of each block is obtained by selecting randomly one of these potential start-times. Then, the block consists of the $L$ oscillation cycles following the chosen start-time, where the random integer $L$ follows a geometric distribution $P(L)\propto (1-q)^{L-1}$, with $0 < q < 1$ and an average block length of $\langle L \rangle = 1/q$ (we have taken $\langle L \rangle = 20$ oscillation cycles, longer than the mean correlation time for all the simulated ``LFPs''). Randomly selected blocks are then concatenated up to the desired length.

When considering a structural motif involving more areas, the ``LFP'' time-series of each area can be resampled jointly or independently. When resampling jointly, matching starting points and block-lengths are selected for each block of the resampled time-series of each area, leading to resampled multivariate time-series in which the structure of causal influences should not be altered. 
The distribution of $\mathrm{TE}_{XY}$ over jointly resampled ``LFP'' time-series describes then for each directed pair of areas $X$ and $Y$ the strength of the corresponding effective connectivity link, as well as the fluctuations of this strength. Conversely, when resampling independently the time-series, start-points and \revs{block-lengths} of the resampled blocks are chosen independently for each area. This second procedure leads by construction to causally independent time-series. Any residual $\widetilde{\mathrm{TE}}$ between directed pairs of independently resampled ``LFPs'' indicates therefore systematic biases, rather than actual causal influences. For each directed pairs of areas $X$ and $Y$, significance of the corresponding causal interaction can be assessed by comparing the bootstrapped distributions of ${\mathrm{TE}}_{XY}(\tau)$ and of $\widetilde{\mathrm{TE}}_{XY}(\tau)$. This comparison is performed in Figures~\ref{fig:EFFECTI_unidir},~\ref{fig:EFFECTI_leaky}~and~\ref{fig:EFFECTI_mutual} and in Supporting~Figure~S3D--E. Here, boxes indicate the median strength of ${\mathrm{TE}}_{XY}(\tau)$ for different directions and the corresponding confidence intervals, comprised between a lower extreme $Q_1-1.5(Q_3-Q_1)$ and and upper extreme $Q_3+1.5(Q_3-Q_1)$, where $Q_1, Q_2$ and $Q_3$ are respectively the first, the second and the third quartiles of the distribution of ${\mathrm{TE}}_{XY}(\tau)$ over jointly resampled time-series. Median values of $\widetilde{\mathrm{TE}}_{XY}(\tau)$ and the corresponding confidence intervals, evaluated as before, are represented by horizontal dashed lines and a surrounding shaded band. When the distributions of $\widetilde{\mathrm{TE}}_{XY}(\tau)$ and $\widetilde{\mathrm{TE}}_{YX}(\tau)$ are not significantly different, a single baseline band is plotted. In this study, strengths and base-line for significancy of effective connectivity for each direction are validated based on, respectively, 500 jointly resampled and 500 independently resampled replicas.

\revs{Note that geometric bootstrap can be applied to arbitrary signals, and does not depend on their strict periodicity. However it is precisely the strong periodic component of our signals that makes necessary the use of geometric bootstrap techniques. Indeed, conventional bootstrap, strongly disrupting signal periodicity, would lead to artificially low thresholds for statistical 
significance of TE (not shown).}

\subsection*{Entropy and Mutual Information (MI)}
We evaluate information transmission between pairs of mono-synaptically connected cells in different areas, linked by a TL-synapse (TL pairs) or by a normally weak long-range synapse (control pairs). Inspired by \cite{Strong1998}, spike trains are digitized into binary streams 
$\mathbf{s}_i(k)$, where $\mathbf{s}_i(k)$ = 1 or 0 respectively when neuron $i$ fires or does not fire during the $k$-th local oscillation cycle (cycle counting is performed independently for each area and includes all the oscillation cycles following a common reference initial time). Note that neurons fire very sparsely and, due to the elevated degree of synchrony in our model, only in narrow temporal intervals centered around the peaks of the ongoing ``LFP'' oscillations. In particular, they fire at maximum once per oscillation cycle. Thus, this oscillatory spiking activity is naturally quantized in time and binning \cite{Strong1998} is not required. For each considered directed pair of cells ($i$ source cell, $j$ target cell), based on very long duration spike trains, we sample normalized histograms for three probability distributions:  $P_i = P(\mathbf{s}_i(k))$, $P_j = P(\mathbf{s}_j(k))$ and $P_{ij} = P(\mathbf{s}_i(k), \mathbf{s}_j(k'))$. When sampling the joint probability distribution $P_{ij}$ we have to distinguish two cases: (i) If the presynaptic cell $i$ belongs to a leader area, i.e. the oscillation of the source area leads in phase over the oscillation of the target area of the considered synapse, then $k' = k$; (ii) Conversely, if the presynaptic cell $i$ belongs to a laggard area, i.e. the oscillation of the target area leads in phase over the oscillation of the source area of the considered synapse, then $k' = k+1$. This means that we seek for spiking correlations only in pairs of spiking (or missed spiking) events in which the ``effect'' follows temporally its potential ``cause'', since physical information transmission cannot occur backward in time. As for the estimation of TE (see previous section), the probabilities $P_i$, $P_j$ and $P_{ij}$ are sampled separately for each specific phase-locking configuration of the ongoing ``LFPs''. Epochs in which the system switches to a different phase-locking configuration, as well as transients following state switchings are dropped. The evaluation of these probabilities is thus based on disconnected spike train chunks, including overall $O(10^4)$ oscillation cycles. Based on these probabilities, the Shannon entropy H of the spike train of the presynaptic neuron $i$ (measuring the information content in its activity) is evaluated as:
\begin{equation}\label{eq:H}
\mathrm{H}_i = -\sum P_i\log_2 P_i
\end{equation}
and MI between pre- and postsynaptic cells as:
\begin{equation}\label{eq:MI}
\mathrm{MI}_{ij} = \sum P_{ij}\log_2 \frac{P_{ij}}{P_i P_j}
\end{equation}
MI is then normalized by the entropy of the pre-synaptic cell, in order to measure the relative efficiency of information transmission along each TL or control synapse.

Statistics are taken over 400 pairs of cells per synapse set, i.e. one set of strong synapses per embedded TL, plus one set of (control) weak synapses. The box-plots in Fig.~\ref{fig:INFOTRANS}C--D report median efficiencies of information transmission efficiencies (for different active effective connectivities), as well as their confidence intervals, estimated non-parametrically from distribution quartiles, as discussed above for TE.
Both MI and H are computed for (finite) spike trains of the largest available length $L$. Following \cite{Strong1998, Panzeri2007}, it is possible to correct these results for finite-size sampling bias (see Supporting~Figure~S4). MI and H are computed again, based on randomly selected shorter matching sections of the full length spike trains. The results of MI / H obtained for various shorter lengths $L / q$ are then plotted against the so-called inverse data fraction $q$, where $q=1$ correspond then to estimations based on full length spike trains. Quadratic extrapolation to $q = 0$ provides a debiased estimation of \mbox{MI / H}. Note that, in order to allow a more direct comparison with the non-debiased TE analysis, the results plotted in Fig.~\ref{fig:INFOTRANS}C--D do not include any finite-size correction. As a matter of fact, as discussed in Supporting~Figure~S4, finite size bias induces a small quantitative overestimation of information transmission efficiency (from $\sim 3\%$ to $\sim 6\%$), that does not affect qualitatively any of the results presented here.

\section*{Acknowledgments}
We thank Albert Compte for useful suggestions, Olav Stetter for figure editing \revs{and the anonymous reviewers for helping us improving our manuscript.}
DB wishes to credit Nicolas Brunel and David Hansel for earlier collaboration on the models used in this study.


\section*{Figure Legends}

\vspace{1em}\noindent
{\bf Figure 1. Flexibility of brain function requires dynamic effective connectivity}. 
This is illustrated by the example of a Giuseppe Arcimboldo's painting ({\it Vertumnus}; 1590, Skoklosters Slott, Sweden). A: the illusion of seeing a face is due to the default activation of a network of brain areas dedicated to face recognition. B: however, selective attention to individual components ---e.g.~to a pear or a flower--- suppresses this illusion by modulating the interaction between these and other brain areas. Therefore, effective connectivity, i.e. the specific active pattern of inter-areal influences, needs to be rewired ``on demand'' in a fast and reliable way, without changes in the underlying structural connectivity between the involved areas.

\vspace{2em}\noindent
{\bf Figure 2. Models of interacting areas}.
 A: in the network model, each local area is modeled as a large network of randomly and sparsely interconnected excitatory and inhibitory spiking neurons (inhibitory cells and synapses are in blue, excitatory cells and synapses are in red, $n_E = n_I = 4000$). Individual neurons spike irregularly (see the spike trains of eight representative neurons, top right), but the activity of the network undergoes a collective fast oscillation, visible in the average membrane potential (see example ``LFP'' trace, bottom right). B: in the rate model, each local area is modeled by a single mean-field rate unit with delayed local inhibition (of strength $K_I < 0$). Its dynamics, describing the average area activity, also undergoes a fast oscillation (see example rate trace, right). C--D: the interaction between multiple local areas ($N = 2$ in the case of the reported graphical illustrations, green and orange shading indicate separate areas) is modeled by the dynamics: of multiple local spiking networks, mutually interconnected by long-range excitatory synapses (see panel C); or of multiple rate units, coupled reciprocally by delayed excitation (of strength $K_E > 0 $, see panel D).

\vspace{2em}\noindent
{\bf Figure 3. Effective motifs of the unidirectional driving family}. 
\revs{For weak inter-areal coupling strengths, out-of-phase lockings of local periodic oscillations give rise to a family of ``unidirectional driving'' effective motif.  The figure shows dynamics and corresponding effective connectivities for fully symmetric structural motifs with $N=2$ (panels A-B) or $N=3$ (panels C-D) areas. A:  the dynamics of $N=2$ interacting areas (green and orange colors) is illustrated by ``LFPs'' (left, top row) and representative spike trains (left, middle row, two cells per each area) from the network model (horizontal bar is 20 ms, vertical bar is 20 mV), as well as by matching rate traces (left, bottom row) from the rate model (arbitrary time units). The right sub-panel reports} the associated effective connectivity measured by Transfer Entropy (TE), evaluated from ``LFPs'' time-series, for all possible directed interactions (indicated by colored arrows). Boxes indicate the interquartile range and whiskers the confidence interval for the estimated TEs. TEs above the grey horizontal band indicate statistically significant causal influences (see {\it Methods}).  \revs{B: to the right of the corresponding box-plot,} effective connectivity is also represented in a diagrammatic form. Arrow thicknesses encode the strength of corresponding causal interactions (if statistically significant). \revs{Below this effective motif, a second motif in the same unidirectional driving family is plotted (with a smaller size), corresponding to} another motif version with equivalent overall topology but \revs{reversed} directionality. 
\revs{The parameters used for $N=2$ are, for} the network model: $p_I = 0.25$, $p_E = 0.01$; \revs{and} for the rate model: $K_I = -250$, $K_E = 5$, $D=\overline{D} = 0.1$.  
\revs{C: this panels reports similar quantities as panel A, but now for a structural motif with $N=3$ areas (green, orange and light blue colors). Effective connectivity is now measured by partialized Transfer Entropy (pTE; see {\it Methods}), in order to account only for direct causal interactions. D: the six effective motifs of the unidirectional driving family for $N=3$ are also reported. The parameters used for $N=3$ are, for} the network model: $p_I = 0.33$, $p_E = 0.006$; \revs{and} for the rate model: $K_I = -300$, $K_E = 5$, $D=\overline{D} = 0.1$.  

\vspace{2em}\noindent
{\bf Figure 4. Effective motifs of the leaky driving family}. 
\revs{The figure shows dynamics and corresponding effective connectivities for fully symmetric structural motifs with $N=2$ (panels A-B) or $N=3$ (panels C-D) areas, for intermediate inter-areal coupling strength, leading to asymmetrically irregular oscillations, phase-locked with an average out-of-phase relation. A:  the dynamics of $N=2$ interacting areas (green and orange colors) is illustrated by ``LFPs'' (left, top row) and representative spike trains (left, middle row, two cells per each area) from the network model (horizontal bar is 20 ms, vertical bar is 20 mV), as well as by matching rate traces (left, bottom row) from the rate model (arbitrary time units). The right sub-panel reports} the associated effective connectivity measured by Transfer Entropy (TE), evaluated from ``LFPs'' time-series, for all possible directed interactions (indicated by colored arrows). Boxes indicate the interquartile range and whiskers the confidence interval for the estimated TEs. TEs above the grey horizontal band indicate statistically significant causal influences (see {\it Methods}).  \revs{B: to the right of the corresponding box-plot,} effective connectivity is also represented in a diagrammatic form. Arrow thicknesses encode the strength of corresponding causal interactions (if statistically significant). \revs{Below this effective motif, a second motif in the same unidirectional driving family is plotted (with a smaller size), corresponding to} another motif version with equivalent overall topology but \revs{reversed} directionality. 
\revs{The parameters used for $N=2$ are, for} the network model: $p_I = 0.25$, $p_E = 0.09$; \revs{and} for the rate model: $K_I = -250$, $K_E = 25$, $D=\overline{D} = 0.1$.  
\revs{C: this panels reports similar quantities as panel A, but now for a structural motif with $N=3$ areas (green, orange and light blue colors). Effective connectivity is measured by partialized Transfer Entropy (pTE; see {\it Methods}), in order to account for direct but not for indirect causal interactions. D: the six effective motifs of the unidirectional driving family for $N=3$ are also reported. The parameters used for $N=3$ are, for} the network model: $p_I = 0.33$, $p_E = 0.06$; \revs{and} for the rate model: $K_I = -300$, $K_E = 11$, $D=\overline{D} = 0.1$.  

\vspace{2em}\noindent
{\bf Figure 5. Effective motifs of the mutual driving family}. 
\revs{The figure shows dynamics and corresponding effective connectivities for fully symmetric structural motifs with $N=2$ (panels A-B) or $N=3$ (panels C-D) areas, for large inter-areal coupling strength, leading to symmetrically irregular oscillations, without a stable phase relation. A:  the dynamics of $N=2$ interacting areas (green and orange colors) is illustrated by ``LFPs'' (left, top row) and representative spike trains (left, middle row, two cells per each area) from the network model (horizontal bar is 20 ms, vertical bar is 20 mV), as well as by matching rate traces (left, bottom row) from the rate model (arbitrary time units). The right sub-panel reports} the associated effective connectivity measured by Transfer Entropy (TE), evaluated from ``LFPs'' time-series, for all possible directed interactions (indicated by colored arrows). Boxes indicate the interquartile range and whiskers the confidence interval for the estimated TEs. TEs above the grey horizontal band indicate statistically significant causal influences (see {\it Methods}).  \revs{B: to the right of the corresponding box-plot,} effective connectivity is also represented in a diagrammatic form. Arrow thicknesses encode the strength of corresponding causal interactions (if statistically significant). \revs{A single motif is included in this family}
\revs{The parameters used for $N=2$ are, for} the network model: $p_I = 0.25$, $p_E = 0.15$; \revs{and} for the rate model: $K_I = -250$, $K_E = 27$, $D=\overline{D} = 0.1$.  
\revs{C: this panels reports similar quantities as panel A, but now for a structural motif with $N=3$ areas (green, orange and light blue colors). Effective connectivity is measured by partialized Transfer Entropy (pTE; see {\it Methods}), in order to account for direct but not for indirect causal interactions. D: the mutual driving effective motif  for $N=3$ is also reported. The parameters used for $N=3$ are, for} the network model: $p_I = 0.33$, $p_E = 0.1$; \revs{and} for the rate model: $K_I = -300$, $K_E = 15$, $D=\overline{D} = 0.1$.

\vspace{2em}\noindent
{\bf Figure 6. Dynamic control of effective connectivity}. 
A: symmetric structural motifs can give rise to asymmetric dynamics in which one area leads in phase over the other (spontaneous symmetry breaking). Basins of attraction (in phase-shift space) of distinct phase-locking configurations are schematically shown here (for $N=2$). Empty circles stand for unstable in- and anti-phase lockings and filled circles for stable out-of-phase lockings (corresponding to unidirectional driving effective motifs). B: phase-shift evolution function $\Gamma(\Delta\phi)$ for the rate model (left, analytical solution, $K_I = -250$) and for the network model (right, numerical evaluation, $p_I = 0.25$). Empty and filled circles denote the same stable and unstable phase-lockings as in panel A. C: cartoon of successful (dashed green arrow) and unsuccessful (dashed grey arrow) switchings induced by brief perturbations (lightning icon). An input pulse to the system destabilizes transiently the current phase-locking (solid red and green arrows). For most perturbations, the system does not leave the current basin of attraction and the previous effective motif is restored (dashed red arrow). However, suitable perturbations can lead the system to switch to a different effective motif (dashed green arrow). D: a pulse of strength $h$ induces a phase advancement of the collective oscillations, depending on its application phase $\phi$, as described by the Phase Response Curve  $Z(\phi)$ (left, rate model; analytical solution, $K_I = -250$) or by the induced shift $\delta\phi(\phi; h)$ (right, network model; numerical evaluation, $p_I = 0.25$). E--F:  frequency histogram of successful switching for pulses applied at different phases (the laggard area is perturbed; $h=0.2I$ for the rate model and $h=500\mbox{ pA}$ for the network model). Predicted intervals for successful switching are marked in green, for the unidirectional (panel E) and for the leaky effective driving (panel F) motifs (left, rate model;  right, network model; parameters as in Figures~\ref{fig:EFFECTI_unidir}~and~\ref{fig:EFFECTI_leaky}). Diagrams of the induced transitions are shown in the third column (see SI, Figure S2 for perturbations applied to the leader area).

\vspace{2em}\noindent
{\bf Figure 7. Effective entrainment}. 
A: examples of rate oscillations for different values of the inter-areal coupling in the rate model ($K_I=-250$, $K_E = 4, 7$ and 8.5, from bottom to top). Filled circles denote peaks of oscillation cycles, different color fillings denote different peak amplitudes. B: The oscillatory dynamics is qualitatively altered by increasing inter-areal coupling, as visualized by bifurcation diagrams, constructed by plotting different peak amplitudes at constant $K_E$, as different dots  (the dots corresponding to the peak amplitudes in panel A, are highlighted also here by filled circles of matching colors). Varying $K_E$ in a continuous range, these dots trace a complex branched structure, denoting emergence of novel dynamical states. 
The bifurcation diagrams for the case of two symmetrically connected areas (top) and two unidirectionally connected areas (bottom) are very similar. For a symmetric structural motif, spontaneous symmetry breaking leads to effective entrainment, mimicking the direct entrainment, which occurs for an asymmetric unidirectional structural motif. Leader and laggard areas in effective entrainment behave similarly to the driver and driven area in direct entrainment (orange and green bifurcation diagrams, respectively). Note that different structural motifs give rise to equivalent effective motifs (see side diagrams). \revs{Note: a different version of panel B was previously published in \cite{Battaglia2007} as Supplementary Figure F.}

\vspace{2em}\noindent
{\bf Figure 8. Effective connectivity affects information propagation}. 
A: in the case of sparsely synchronized oscillations, individual neurons fire irregularly (see four example spike trains, middle row) even when the local area activity undergoes a very regular collective rhythm (evident in ``LFP'' traces, bottom row). Therefore, a large amount of information can be potentially encoded, at every (analog) oscillation cycle, in the form of (digital-like) codewords in which ``1'' or ``0'' entries denote respectively firing or missed firing of a specific neuron in the considered cycle (top row). B: the strength of specific subsets of long-range excitatory synapses is systematically enhanced in order to form unidirectional ``transmission lines'' (TLs) embedded into the $N=2$ symmetric structural motif (see {\it Methods}). Cells and synapses belonging to TLs are highlighted by pale green (``green-to-orange'' area direction) and lilac (``orange-to-green'' area direction) colors. Communication efficiency along TLs is quantified by the Mutual Information ($\mathrm{MI}$) between spike trains of pairs of source and target cells connected directly by a TL synapse, normalized by the entropy ($\mathrm{H}$) of the source cell. C--D: boxplots (see Figures~\ref{fig:EFFECTI_unidir},~\ref{fig:EFFECTI_leaky}~and~\ref{fig:EFFECTI_mutual}) of $MI/H$ for different groups of interconnected cells and for different active effective motifs. Pale green and lilac arrows below the boxplots indicate pairs of cells interconnected by the TL marked with the corresponding color. A dot indicates control pairs of cells interconnected by ordinary weak long-range synapses. Green and orange arrows indicate the dominant directionality of the active effective connectivity motif.
C: unidirectional driving effective motif family. Communication efficiency is enhanced only along the TL aligned to the directionality of the active effective connectivity, while it is undistinguishable from control along the other TL.  D: leaky driving effective motif family. Communication efficiency is enhanced along both TLs, but more along the TL aligned to the dominant directionality of the active effective connectivity.

\vspace{2em}\noindent
\textbf{Figure 9.Transfer entropy depends on time lag and quantization}. \revs{A--C: The matrices in these panels illustrate the dependence of TE (network model, $N=2$ fully symmetric structural motif, cfr. Figures~\ref{fig:EFFECTI_unidir},~\ref{fig:EFFECTI_leaky}~and~\ref{fig:EFFECTI_mutual}) on the number $B$ of discretization bins used to describe the time-series of neural activity and on the adopted time lag $\tau_{lag}$ between the time-series (see \textit{Methods}). The matrices in the first two columns (from the left) report TEs in the two possible interaction directions, $\mathrm{TE}_{XY}$ and $\mathrm{TE}_{YX}$, and the matrices in the third column visualize the causal unbalancing $\Delta\mathrm{TE}$ ($-1 \le \Delta\mathrm{TE}\le 1$), which quantifies the asymmetry between causal influences in the two directions (see {\it Methods}). All of these quantities are evaluated for different combinations of $B$ and $\tau_{lag}$. The vertical axes of the matrices correspond to the range $2 < B < 200$ bins and the horizontal axes to the range 1 ms $ < \tau_{lag} <$ 60 ms. This range of time lags corresponds approximately to three oscillation periods. Horizontal scale lines indicate the average oscillation period ($\langle T\rangle =$ 16.4, 18.9 and 19.1 ms, respectively for panels A, B and C). Values of TE and $\Delta\mathrm{TE}$ are color-coded (see color bars at the bottom, note the two different color scales for TE and $\Delta\mathrm{TE}$). Black dotted lines in the matrices enclose regions in which $\mathrm{TE}_{XY}$ or $\mathrm{TE}_{YX}$ rise above the threshold for significancy of the corresponding causal interaction (see {\it Methods}). These significance contours are overlayed in the corresponding  $\Delta\mathrm{TE}$ matrix. A star denotes the combination of $B$ and $\tau_{lag}$ used for the analysis throughout the main article ($\tau_{\mbox{{\tiny lag}}}= 5$ ms, $B = 175$). Different rows report TE matrices for different effective motifs. A: unidirectional driving effective motif. B: leaky driving effective motif. C: mutual driving effective motif. Diagrams of these effective motifs are drawn in the fourth column as a visual reference. All other parameters are as for the analyses of Figures~\ref{fig:EFFECTI_unidir},~\ref{fig:EFFECTI_leaky}~and~\ref{fig:EFFECTI_mutual}.}

\section*{Supporting Information Legends}

\vspace{1em}\noindent
\textbf{Supporting Figure S1. Phase reduction and phase response}. A: oscillating time-series (in the example, a ``LFP'' time-series from the network model) can be described in terms of phase, even if they are not periodic in strict sense, by interpolating linearly an instantaneous empiric phase variable $\phi$ to the oscillation cycles (generally of unequal lengths). B: the application of a pulse current $\delta I$ induces a shift $\delta\phi$ in the oscillation phase of the ongoing oscillation (in the example, a rate trace from the rate model). The amplitude of the induced shift depends on the phase $\phi$ of the ongoing oscillation at which the perturbation is applied. 

\vspace{1em}\noindent
\textbf{Supporting Figure S2. Dynamic control of effective connectivity (perturbation applied to the leader area)}. 
A--B: frequency histogram of successful switching for pulses applied at different phases ($h=0.2I$ for the rate model and $h=500\mbox{ pA}$ for the network model). Predicted intervals for successful switching are marked in green, for the unidirectional (panel E) and for the leaky effective driving (panel F) motifs (left, rate model;  right, network model; parameters as in Figures~\ref{fig:EFFECTI_unidir}~and~\ref{fig:EFFECTI_leaky}). Diagrams of the induced transitions are shown in the third column (see Figure~\ref{fig:PRC} for perturbations applied to the laggard area).

\vspace{1em}\noindent
\textbf{Supporting Figure S3.  Effective connectivity with transmission lines (TLs).}  We consider a fully symmetric structural motif of $N=2$ structurally connected areas with embedded unidirectional TLs. 
\revs{Synapses involved in TLs are enhanced by multiplying the ordinary excitatory peak conductance by a multiplier $K_{TL}$. Raster plots relative to the spiking activity of excitatory neurons of the two areas are shown in panels A--C (green and orange color denote spikes of excitatory neurons from different populations, the horizontal scale line corresponds to 20 ms) for a weak inter-areal coupling (unidirectional driving effective motif, see Figure~\ref{fig:EFFECTI_unidir} for parameters). A: when $K_{TL}=0$ (no TL embedded), the synchronous oscillations of the two populations lock in an out-of-phase fashion. B: for $K_{TL}=22$ (just below a critical value), the raster plot of the spiking activity is virtually indistinguishable from the raster plot of panel A. C: for $K_{TL} = 22.5$ (just above a critical value), the oscillations of the two populations lock in an in-phase configuration.
D--E:} Effective connectivities associated to different dynamical states are measured by Transfer Entropy (TE), evaluated from ``LFPs'' time-series, for all possible directed interactions (indicated by green or orange arrows). Boxes indicate the interquartile range and whiskers the confidence interval for the estimated TEs. TEs above the grey horizontal band indicate statistically significant causal influences (see {\it Methods}).  In each plot, the third and the fourth boxes (from left to right) refer to TEs evaluated from ``LFPs'' restricted to groups of neurons that are source and target of a TL (pale green color denotes TL in the ``green-to-orange'' area direction, lilac color denotes TL in the ``orange-to-green'' area direction). Below each TE box-plot, effective connectivity is also represented in a diagrammatic form. Arrow thicknesses encode the strength of corresponding causal interactions (if statistically significant).  \revs{D: TEs for the unidirectional driving effective motif with embedded TLs ($K_{TL} = 22$). E: TEs for the leaky driving effective motif with embedded TLs ($K_{TL} = 24.5$).} Comparing these effective motifs with Figures~\ref{fig:EFFECTI_unidir}~and~\ref{fig:EFFECTI_leaky}, we conclude that the embedding of TLs does not alter the overall effective connectivity.

\vspace{1em}\noindent
\textbf{Supporting Figure S4.  Scaling of Mutual Information (MI) with spike train length}. 
MI normalized by entropy (at optimal time lag) is plotted against the inverse data fraction $q$. For each data fraction $q$, several
bivariate spike trains are extracted from the original long spike trains (3 min, $q = 1$) and the mean MI
is further averaged over these reduced-length spike trains. Asymptotic values are extrapolated
through a quadratic interpolation. Error bars correspond to standard error. A: unidirectional driving effective motif, 
MI along the TL in the leader-to-laggard direction (pale green color), extrapolated asymptotic value is $\mathrm{MI} / \mathrm{H} = 0.683$. 
B: unidirectional driving effective motif, MI along the TL in the laggard-to-leader direction (lilac color), extrapolated asymptotic value is $\mathrm{MI} / \mathrm{H} = 0.0066$.
In both cases, the finite size of the used spike trains produces a positive but small bias in the estimation of MI. Compared to Figure~\ref{fig:INFOTRANS}C, for the leader-to-laggard direction the overestimation is of $\sim3\%$ and for the laggard-to-leader direction is of $\sim 6\%$. 

\vspace{1em}\noindent
\textbf{Supporting Text S1.  Full description of model parameters and complete analytic expressions.} This text contains the following sections: i) Model neurons; ii) Model synapses; iii) Parameters of the background noise; iv) Phase response of the rate model; v) Phase-locking in the rate model.

\newpage

\begin{flushleft}
{\Large
\textbf{Supporting Text S1}
}
\end{flushleft}

\section*{Model neurons} 
We use the Wang-Buzsáki (WB) conductance-based model (ref.  \textbf{[86]} in main text) to describe each single excitatory and inhibitory neuron. The WB model is described by a single compartment endowed with sodium and potassium currents. The membrane potential is given by:
\begin{equation*}
C\frac{dV}{dt} = -I_L -I_{Na} -I_{K} +I_{ext} +I_{rec}
\end{equation*}
where $C$ is the capacitance of the neuron, $I_L = g_L (V-V_L )$ is the leakage current, $I_{ext}$ is an external driving current and $I_{rec}$ is due to recurrent interactions with other neurons in the network (see later). Sodium and potassium currents are voltage-dependent and given by $I_{Na} =g_{Na}m^3_{\infty}h(V-V_{Na})$ and $I_K =g_{K}n^4(V-V_{K})$. The activation of the sodium current is instantaneous: 
\begin{equation*}
m_\infty(V) = \frac{\alpha_m(V)}{\alpha_m(V)+\beta_m(V)}
\end{equation*}
Sodium current inactivation and potassium current activation evolve according to:
\begin{equation*}
\frac{dx}{dt}=\Phi\cdot\left(\alpha_x(V)(1-x)-\beta_x(V)x\right)
\end{equation*}
where $x=h,n$ and $\alpha_x$ and $\beta_x(V)$ are non-linear functions of the membrane potential given by:
\begin{eqnarray*}
\alpha_m(V) &=& \frac{0.1\/(V+35)}{1+e^{-\frac{V+35}{10}}}\\
\beta_m(V) &=& 4\/e^{-\frac{V+60}{18}}\\
\alpha_n(V) &=& \frac{0.03\/(V+34)}{1-e^{-\frac{V+34}{10}}}\\
\beta_n(V) &=& 0.375\/e^{-\frac{V+44}{80}}\\
\alpha_h(V) &=& 0.21\/e^{-\frac{V+58}{20}}\\
\beta_h(V) &=& \frac{3}{1+e^{-\frac{V+28}{10}}}
\end{eqnarray*}
Other parameters are $g_{Na} = 35$ mS/cm$^2$, $V_{Na}=55$ mV, $g_K=9$ ms/cm$^2$, $V_K=-90$ mV, $g_L=0.1$ mS/cm$^2$, $C=1\,\,\mu$F/cm$^2$ and $\phi=5$.

\section*{Model synapses}
The synaptic current induced in a postsynaptic neuron by a single presynaptic action potential is given by $I_{spike}(t) = -g_{x}\/s_{spike}(t)\/(V-V_{x})$, where $V$ is the potential in the postsynaptic neuron and $V_x$ is the reversal potential of the synapse (for excitatory synapses $V_E = 0$ mV, for inhibitory synapses $V_I = -80$ mV). The time-course of the postsynaptic conductance is described by:
\begin{equation*}
s_{spike}(t) \propto \/\left(\exp\left(-(t+d-t^*)/\tau_1\right)-\exp\left(-(t+d-t^*)/\tau_2\right)\right)
\end{equation*}
for $t>t^*$, 0 otherwise, where $t^*$ is the time of the presynaptic spike, $d$ is the latency, $\tau_1$ the rise-time and $\tau_2$ the decay-time. The total recurrent current $I_{rec}(t)$ is the sum of time-dependent contributions $I_{spike}(t)$ from all the presynaptic spikes fired to time $t$. 
The normalization constant of $s_{spike}(t)$ is chosen such as the peak value of $s_{spike}$ is equal to 1. 
For all simulations in the paper, we take $\tau_1=1$ ms, $\tau_2 = 3$ ms and $d = 0.5$ ms. Thus, post-synaptic currents have a relatively fast decay, corresponding to AMPA-like excitatory and GABA$_A$-like inhibitory synapses.
For simplicity, we take only two possible peak conductances, $g_I = 90$ $\mu$S/cm$^2$ for inhibitory synapses within an area and $g_E = 5$ $\mu$S/cm$^2$ for excitatory synapses within and between areas.

\section*{Parameters of the background noise}
In addition to recurrent synaptic inputs, each neuron receives a noisy input, representing background spiking activity. It is modeled as an excitatory current having the same functional form of a recurrent current induced by a Poisson spike train with firing rate $f_{ext}$. The peak conductance of this noisy background input is $g_{ext}$. In our simulations, we take 
$f_{ext} = 5$ kHz, and $g_{ext} = g_E = 5$ $\mu$S/cm$^2$. Each neuron is driven by statistically independent Poisson noise realizations.

\section*{Phase response of the rate model}
As previously throughly reported in the Supplementary Material of ref.  \textbf{[42]} (in main text), 
the firing rate of a single oscillating area (only local inhibitory coupling $K_I < 0$ with delay $D$) can be derived analytically assuming that: (i) the total input current $I_{tot}(t) = I + K_IR(t-D)$ is below threshold (i.e. negative) for a duration $T_{st} > D$; (ii) the delay $D$ and the oscillation period $T$ fulfill the inequalities $D < T-T_{st} < 2D$. The
conditions (i) and (ii) hold for sufficiently strong local inhibition, and, specifically, for the value $K_I= -250$ and the delay $D = 0.1$ adopted in the main paper. Under these conditions, the limit cycle of the firing rate assumes then the following analytic form (see Figure 2B in the main paper):
\newpage
\begin{equation*}
R(t) = R_{peak} \cdot \left\{\begin{array}{ll}
e^{-t} & t\in[0,T_{st}] \\
e^{-t}+K_Ie^D\left[e^{-t}-e^{-T_{st}}+e^{-t}(t-T_{st})\right] & t\in[T_{st},T_{st}+D] \\
e^{-t}+K_Ie^D\left[e^{-t}-e^{-T_{st}}+e^{-t}(t-T_{st})\right]  & \\
\quad + K_I^2e^{2D}\left[e^{-t}-e^{-D-T_{st}}+e^{-t}(t-T_{st}-D+\frac{(t-T_{st}-D)^2}{2})\right]& t\in[T_{st}+D,T] 
\end{array}\right.
\end{equation*}
where $R_{peak}$ is the peak amplitude of the periodic oscillation of the rate and depends linearly on the level of the background current $I$. The oscillation period $T$ and the sub-threshold time can be determined numerically by solving the system of non-linear equations:
\begin{eqnarray*}
e^{T-T_{st}}  & = & 1+K_I e^D \left(1+{T-T_{st}}-D-e^{{T-T_{st}}-D}\right) \\
e^T \quad &  =  &1+K_I e^D (1-e^{T-T_{st}}+{T-T_{st}})+K_I^2 e^{2D} \left(1-e^{{T-T_{st}}-D}+{T-T_{st}}-D+\frac{({T-T_{st}}-D)^2}{2}\right) 
\end{eqnarray*}
We define the phase relative to the oscillation as $\phi(t) = \mod(t-t_0, T)$, where the time-shift $t_0$ is chosen such as 
$\phi(t_{t_peak})=0$ in correspondence of the timings $t_{peak}$ of oscillation peaks. Phases are therefore, with this notation, bounded between 0 and 1. We use this convention throughout all analytic developments for the sake of simplicity. In The results involving phases in the main article are then translated back into the more usual angular range comprised between 0$^\circ$ and 360$^\circ$. The application of a pulse current $\delta I = h \delta(\phi - \phi_p)$ at a phase $\phi_p$ induces a phase-shift $\delta\phi(\phi_p) = hZ(\phi_p)$ (see Figure S3B). The analytic expression for the Phase Response Curve (PRC) $Z(\phi)$ can be derived from the knowledge of the limit cycle solution, and reads:
\begin{equation*}
Z(\phi) = R_{peak}\cdot\left\{\begin{array}{ll}
0 & \phi\in\left[0,\phi_{st}\right] \\
-e^{T(\phi-1)}\left(1+K_I T e^D(1-\phi-\phi_D)\right) & \phi\in\left[\phi_{st},1-\phi_D\right] \\
-e^{T(\phi-1)}& \phi\in\left[1-\phi_D,1\right] 
\end{array}\right.
\label{Z}
\end{equation*}
where $\phi_{st} = \frac{T_{st}}{T}$ and $\phi_{D} = \frac{D}{T}$. The resulting PRC is therefore null over a very large interval of phases, leading in this broad range to refractoriness toward perturbations. A plot of $Z(\phi)$ for the parameters used in our study is reported in Figure 4D (main text).

\section*{Phase-locking in the rate model}
As discussed in the main text, the time-evolution of the instantaneous phase shift $\Delta\phi(t)$ between two coupled areas can be described, in the weak coupling limit, by the equation:
\begin{equation*}
\frac{d\Delta \phi}{dt}=\Gamma(\Delta \phi)
\end{equation*}
The term $\Gamma(\Delta\phi)$ is a functional of the phase response and of the limit cycle waveform of the uncoupled oscillating areas. In terms of the previously derived analytic expressions of $Z(\phi)$ and of the rate oscillation limit cycle $R(\phi)$ (phase-reduced) for $K_E = 0$, this functional  can be expressed as $\Gamma(\Delta\phi) = C(\Delta\phi) - C(-\Delta\phi)$, where:
\begin{equation*}
C(\Delta\phi) =  \int_{0}^{1} Z(\phi) R(\phi + \Delta\phi - D)d\phi
\end{equation*}
Stable phase-lockings are therefore given by the zeroes of $\Gamma$ with negative slope crossing.
Analytic expressions for the integral $C(\Delta\phi)$ have already been derived and published in the Supplementary Material of ref.  \textbf{[31]} (in main text). We report here these expression again, in order to make the presentation of results self-contained. To compute $C(\Delta\phi)$, six different intervals of $\Delta \phi$ need
to be considered separately. The result is:
\begin{equation*}
C(\Delta \phi) = \left\{\begin{array}{ll}
C_{00}(\phi_{st},1)+C_{10}(\phi_{st},1-\phi_D) & \Delta \phi \in [\phi_D-\phi_{st},  \phi_{st}+\phi_D-1] \\
\,\\
C_{00}(\phi_{st},1)+C_{10}(\phi_{st},1-\phi_D) +C_{01}(\phi_{st}+\phi_D-\Delta \phi,1) & \Delta \phi \in [\phi_{st}+\phi_D-1,\phi_{st}+2\phi_{D}-1] \\
\,\\
C_{00}(\phi_{st},1)+C_{10}(\phi_{st},1-\phi_D) +C_{01}(\phi_{st}+\phi_D-\Delta \phi,1) & \\
\quad +C_{11}(\phi_{st}+\phi_D-\Delta \phi,1-\phi_D)+C_{02}(\phi_{st}+2\phi_D-\Delta \phi,1)& \Delta \phi \in[\phi_{st}+2\phi_D-1,\phi_D] \\
\,\\
C_{00}(\phi_{st},1+\phi_D-\Delta\phi)+e^T C_{00}(\phi_{st}+\phi_D-\Delta \phi,1) & \\ 
\quad+ C_{10}(\phi_{st},1-\phi_D)+C_{01}(\phi_{st},1+\phi_D-\Delta \phi) & \\
\quad +C_{11}(\phi_{st},1-\phi_D) +C_{02}(\phi_{st}+2\phi_D-\Delta \phi,1+\phi_D-\Delta \phi) & \Delta \phi \in[\phi_D,\phi_{st}-1+3\phi_D] \\
\,\\
C_{00}(\phi_{st},1+\phi_D-\Delta\phi)+e^T C_{00}(\phi_{st}+\phi_D-\Delta \phi,1) & \\
\quad + C_{10}(\phi_{st},1-\phi_D)+C_{01}(\phi_{st},1+\phi_D-\Delta \phi) & \\
\quad +C_{11}(\phi_{st},1-\phi_D)+C_{02}(\phi_{st}+2\phi_D-\Delta \phi,1+\phi_D-\Delta \phi) & \\
\quad + C_{12}(\phi_{st}+2\phi_D-\Delta\phi,1-\phi_D) & \Delta \phi \in[\phi_{st}-1+3\phi_D,2\phi_D] \\
\,\\
C_{00}(\phi_{st},1+\phi_D-\Delta\phi)+e^T C_{00}(\phi_{st}+\phi_D-\Delta \phi,1) & \\
\quad+ C_{10}(\phi_{st},1+\phi_D-\Delta \phi) +e^T C_{10}(1+\phi_D-\Delta \phi,1-\phi_D) & \\
\quad +C_{01}(\phi_{st},1+\phi_D-\Delta \phi)+C_{11}(\phi_{st},1+\phi_D-\Delta \phi)  & \\
\quad +C_{02}(\phi_{st},1+\phi_D-\Delta \phi)+ C_{12}(\phi_{st},1+\phi_D-\Delta \phi) & \Delta \phi > 2\phi_D 
\end{array}\right.
\label{C}
\end{equation*}
where
\begin{eqnarray}
C_{00}(a,b) & = & -(b-a)Te^{-T(1-\Delta\phi+\phi_D)} \nonumber\\
C_{10}(a,b) & = & K_I e^{T(2\phi_D-1-\Delta\phi)}\left[\frac{T(x+\phi_D-1)^2}{2}\right]^b_a\nonumber\\
C_{01}(a,b) & = & -K_I e^{D-T}\left[T(b-a)e^{T(\phi_D-\Delta\phi)}-e^{-T_{st}}(e^{bT}-e^{aT})+e^{T(\phi_D-\Delta\phi)}\left[\frac{T(x+\Delta\phi-\phi_D-\phi_{st})^2}{2}\right]^b_a\right]\nonumber\\
C_{02}(a,b) & = &  -K_I^2 e^{2D-T}\left[(b-a)e^{T(\phi_D-\Delta\phi)}-e^{-D-T_1}(e^{bT}-e^{aT})+ \right. \nonumber\\ 
& & \quad \left.+e^{T(\phi_D-\Delta\phi)}\left[\frac{T(x+\Delta\phi-2\phi_D-\phi_{st})^2}{2}+\frac{T(x+\Delta\phi-2\phi_D-\phi_{st})^3}{6}\right]^b_a\right]\nonumber\\
C_{11}(a,b) & = & K_I^2 e^{2D-T}\left[e^{T(\phi_D-\Delta\phi)}\left(\frac{(bT)^2}{2}-\frac{(aT)^2}{2}+T(D-T)(b-a)\right)\right.-
\\
& &\left.-e^{-T_{st}}\left[(xT-1)e^{xT}+(D-T)e^{xT}\right]^b_a + \right. \nonumber\\ 
& & \quad \left.+e^{T(\phi_D-\Delta\phi)}\left[\frac{(xT+D-T)^3}{3}+T(x+\Delta\phi-2\phi_D-\phi_{st})\frac{(xT+D-T)^2}{2}\right]^b_a\right]\nonumber\\
C_{12}(a,b) & = & K_I^3 e^{3D-T}\left[e^{T(\phi_D-\Delta\phi)}\left(\frac{(bT)^2}{2}-\frac{(aT)^2}{2}+T(D-T)(b-a)\right)\right.-\\
& &\left.-e^{-D-T_{st}}\left[(xT-1)e^{xT}+(D-T)e^{xT}\right]^b_a + \right. \nonumber\\ 
& & \quad \left.+e^{T(\phi_D-\Delta\phi)}\left[\frac{(xT+D-T)^3}{3}+T(1+\Delta\phi-3\phi_D-\phi_{st})\frac{(xT+D-T)^2}{2}+\frac{(xT+D-T)^4}{8}+ \right.\right. \nonumber\\ 
& & \quad \left.\left.+T(1+\Delta\phi-3\phi_D-\phi_{st})\frac{(xT+D-T)^3}{3}+T(1+\Delta\phi-3\phi_D-\phi_{st})^2\frac{(xT+D-T)^2}{4}\right]^b_a\right]\nonumber
\end{eqnarray}
where $[f(x)]^b_a = f(b)-f(a)$. A plot of $\Gamma(\Delta\phi)$ for the parameters used in our study is reported in Figure 4B (main text).

\newpage

\begin{flushleft}
{\Large
\textbf{Figures}
}
\end{flushleft}

\vspace{6em}


\begin{figure}[!ht]
\begin{center}
\includegraphics[width=8.3cm]{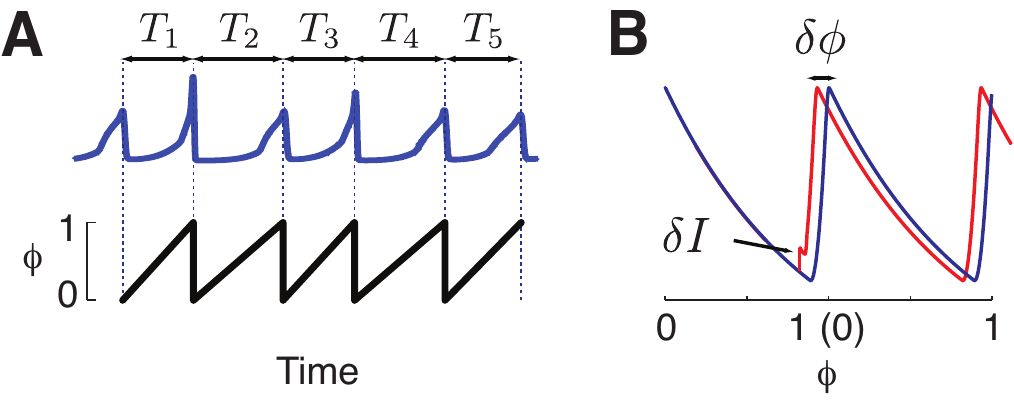}
\end{center}
\caption{ \textbf{[FIGURE S1]}}
\end{figure}

\begin{figure}[!ht]
\begin{center}
\includegraphics[width=12.35cm]{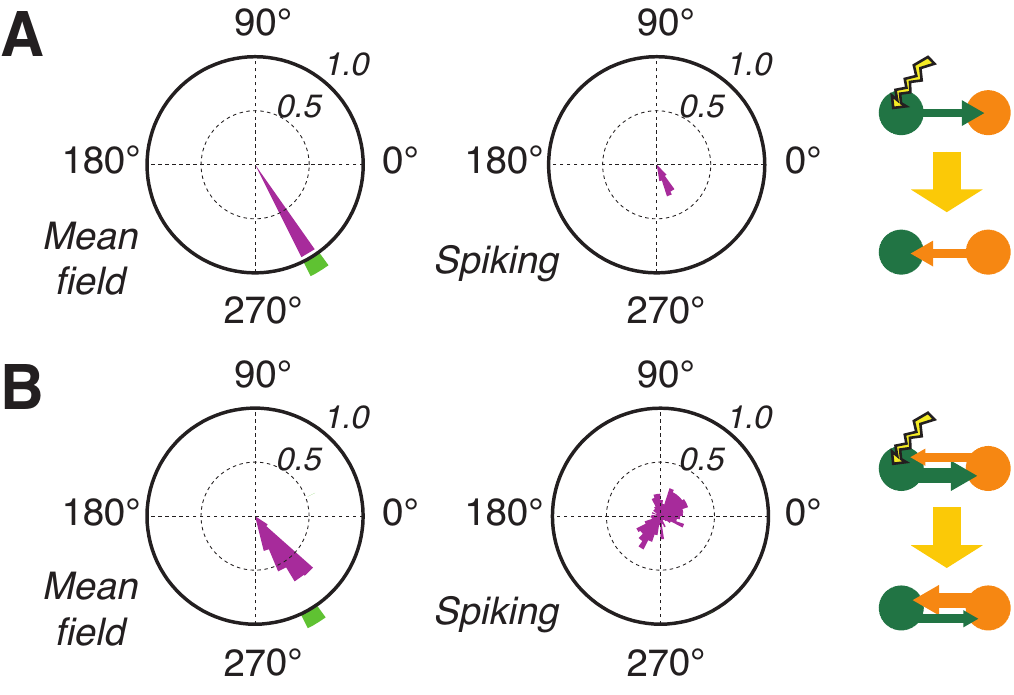}
\end{center}
\caption{ \textbf{[FIGURE S2]}}
\end{figure}

\begin{figure}[!ht]
\begin{center}
\includegraphics[width=12.35cm]{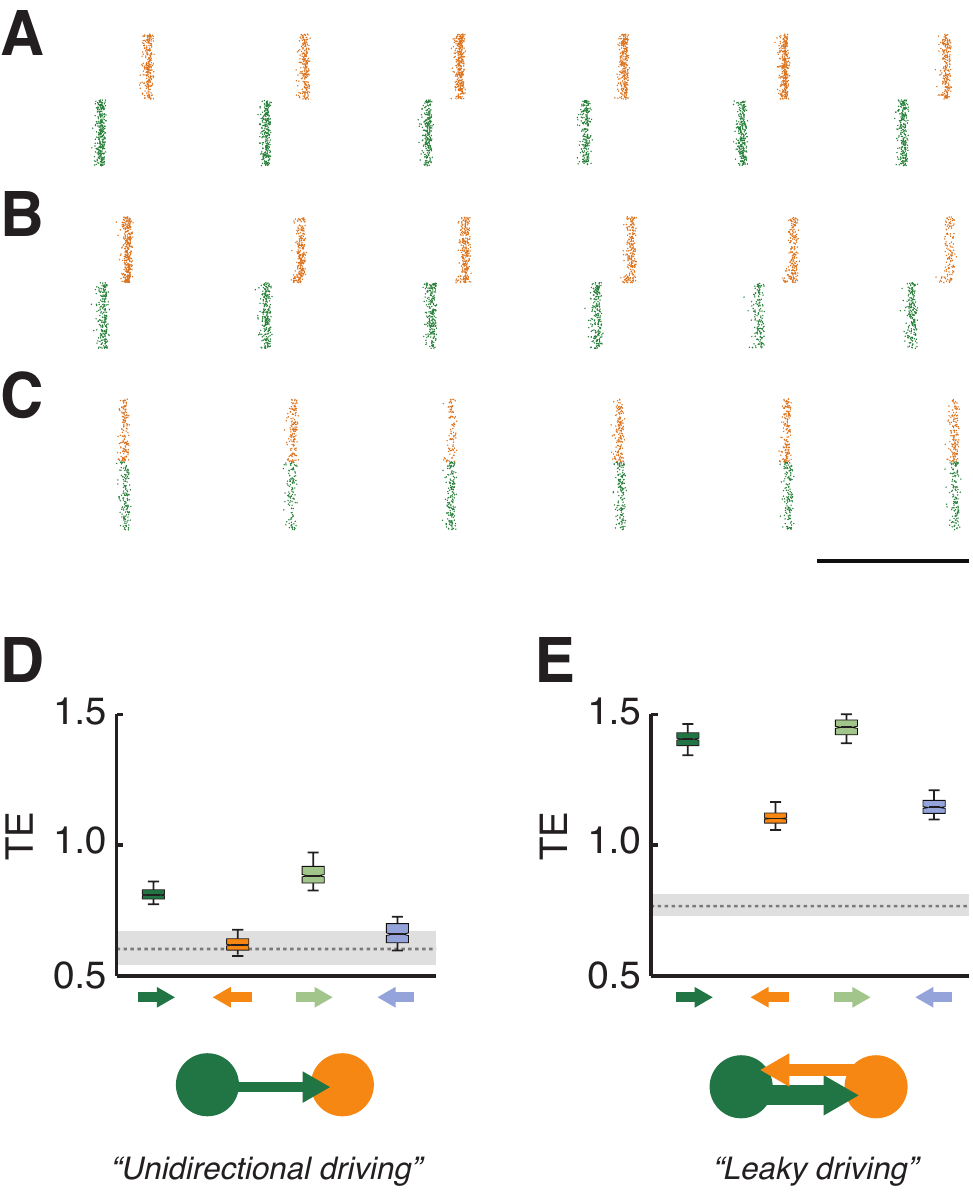}
\end{center}
\caption{ \textbf{[FIGURE S3]}}
\end{figure}

\begin{figure}[!ht]
\begin{center}
\includegraphics[width=12.35cm]{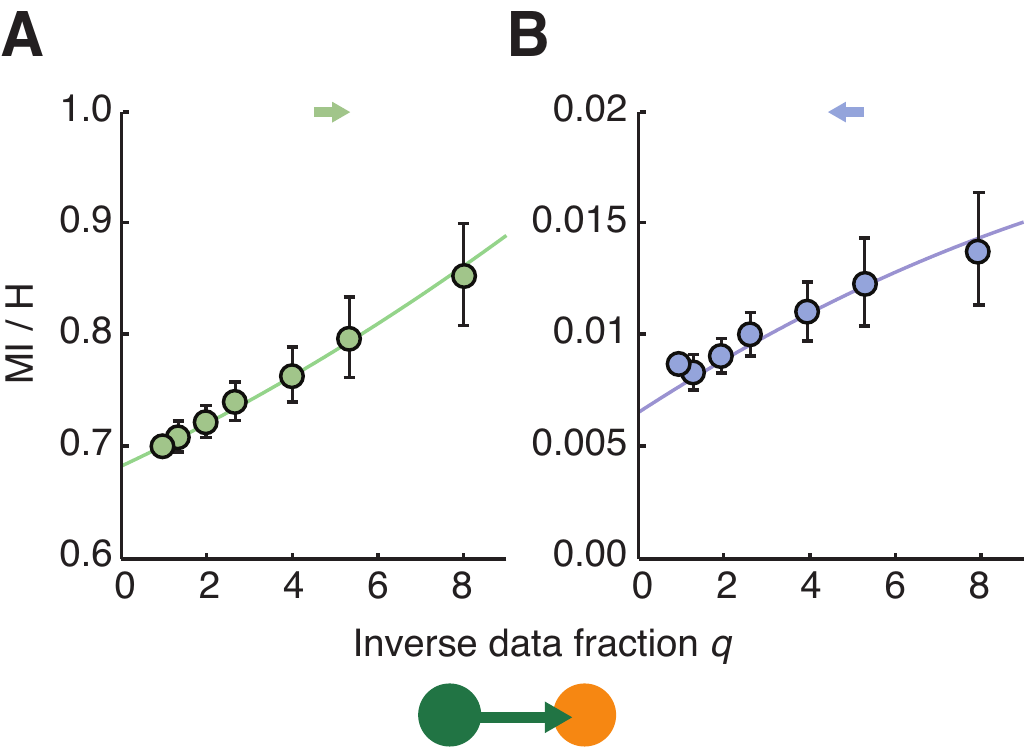}
\end{center}
\caption{ \textbf{[FIGURE S4]}}
\end{figure}

\end{document}